\documentclass[11pt,twoside,letterpaper]{article} %% The same for {book}
\usepackage{times,fancyhdr}
\usepackage[dvips]{graphicx}
\usepackage{hyperref}
\usepackage{fancyheadings,lastpage}
\usepackage{color}
\usepackage{amssymb}
\usepackage{multirow}
\usepackage{amscd}

\usepackage{epsfig}
\usepackage{amssymb}
\usepackage{amsmath}
\usepackage{amsfonts}
\usepackage{amsthm,amscd}
\usepackage{amsbsy}
\usepackage{multirow}  %hhhhhhh
\usepackage{multicol}  %hhhhhhh
\usepackage{latexsym}
\usepackage{bm}
\usepackage{url} 			%% Nicely format and linebreak URLs in the bibliography (and elsewhere).
\usepackage{layout}
\usepackage{pslatex}
\usepackage{cite}
\usepackage{fleqn} 		% displayed formulas flush left (default is centered).
\usepackage{makeidx}
\makeindex 						% Creates index at the end of the book (with makeidx).
\usepackage{layout} 	% To see the current values of these dimensions, use the layout package, 
\usepackage{epstopdf}	% Automatically converts EPS files to encapsulated PDF files (using ghostscript).
\usepackage{color}
\usepackage{hyperref}
\usepackage[latin1]{inputenc}
\usepackage[T1]{fontenc}
\usepackage[letter,cam,center]{crop} %% Printing crop-marks
\usepackage{type1cm} 	%% Use scalable, PostScript Type 1 versions of the Computer Modern fonts.
\usepackage{courier}	%% Replace the standard Computer Modern Typewriter font LaTeX uses
\usepackage{lscape} 	% for landscape section

\makeatletter 
\def\cleardoublepage{\clearpage\if@twoside \ifodd\c@page\else% 
    \hbox{}% 
    \thispagestyle{empty}%
    \newpage% 
    \if@twocolumn\hbox{}\newpage\fi\fi\fi} 
\makeatother

\def\figurename{Figure}
\makeatletter
\renewcommand{\fnum@figure}[1]{\figurename~\thefigure.}
\makeatother

\def\tablename{Table}
\makeatletter
\renewcommand{\fnum@table}[1]{\bfseries\tablename~\thetable.}
\makeatother

\definecolor{cream}{rgb}{.97, .95, .88}
\definecolor{darkcream}{rgb}{1., .88, .5}
\definecolor{lightpink}{rgb}{.98, .88, .87}
\definecolor{lightwhite}{rgb}{1., .98, .95}
\definecolor{lightsalmon}{rgb}{1., .95, .90}
\definecolor{lightviolet}{rgb}{.99, .96, 1.}
\definecolor{lightgray}{rgb}{.96, .96, .96}
\definecolor{lgray}{rgb}{.75, .75, .75}
\definecolor{LemonChiffon}{rgb}{1., 0.98, 0.8}
\definecolor{lightolivegreen}{rgb}{.84, .89, .25}
\definecolor{lightgreen}{rgb}{.664, 1., .52}
\definecolor{llgreen}{rgb}{.900, .983, .960}
\definecolor{tristle}{rgb}{.792, .609, .698}
\definecolor{lighttristle}{rgb}{.892, .709, .798}
\definecolor{pink}{rgb}{.95, .35, .7}
\definecolor{magenta}{rgb}{1., 0, 1.}
\definecolor{violetl}{rgb}{0.820, 0.730, 0.969}
\definecolor{violet}{rgb}{.65, .30, .7}
\definecolor{darkolivegreen}{rgb}{.55, .65, .35}
\definecolor{maroon}{rgb}{.7, .26, .56}
\definecolor{mediumorchid}{rgb}{.8, .33, .83}
\definecolor{mediumorchidd}{rgb}{1., .33, .63}
\definecolor{darkgreen}{rgb}{0.1, .6, .13}
\definecolor{ddarkgreen}{rgb}{0.1, .45, .13}
\definecolor{lightyellow}{rgb}{1., 1., .82}
\definecolor{turquoise}{rgb}{.3, .70, .61}
\definecolor{turquoisel}{rgb}{.66, .94, .83}
\definecolor{darkturquoise}{rgb}{.21, .55, .50}
\definecolor{coral}{rgb}{1., .6, .21}
\definecolor{lightorange}{rgb}{1., .86, 0.69}
\definecolor{orangered}{rgb}{1., .5, 0.}
\definecolor{orange}{rgb}{1., .65, .1}
\definecolor{orangel}{rgb}{1., .85, .3}
\definecolor{darkorange}{rgb}{.875, .4, .204}
\definecolor{ddarkorange}{rgb}{.675, .318, .05}
\definecolor{bluesky}{rgb}{.48, .53, 1.}
\definecolor{gold}{rgb}{1., .85, 0.25}
\definecolor{goldd}{rgb}{.95, .75, 0.05}
\definecolor{darkgold}{rgb}{.85, .65, 0.05}
\definecolor{darkviolet}{rgb}{.54, .04, .84}
\definecolor{ddarkviolet}{rgb}{.382, .063, .657}
\definecolor{ddarkgreen}{rgb}{0.1, .45, .13}
\definecolor{lightblue}{rgb}{.30, .86, .89}
\definecolor{lblue}{rgb}{0.88, 0.90, 0.95}
\definecolor{LightBlue}{rgb}{0.68, 0.85, 0.9}
\definecolor{darkblue}{rgb}{.105, .308, .707}
\definecolor{lightmaroon}{rgb}{0.85, 0.38, 0.58}
\definecolor{darkmaroon}{rgb}{0.604, 0.169, 0.451}
\definecolor{darkpink}{rgb}{0.879, 0.020, 0.766}
\definecolor{ddarkpink}{rgb}{0.738, 0.195, 0.406}
\definecolor{grey}{rgb}{0.717, 0.717, 0.717}
\definecolor{brown}{rgb}{0.640, 0.323, 0.182}
\definecolor{darkbrown}{rgb}{0.34, 0.25, 0.05}
\definecolor{orangebrown}{rgb}{0.433, 0.262, 0.06}
\definecolor{pinkl}{rgb}{1., 0.788, 0.918}
\definecolor{salmon}{rgb}{1., 0.66, 0.5}
\definecolor{lightbrown}{rgb}{0.703, 0.508, 0.121}

\def\Journal#1#2#3#4{{#1} {\bf #2}, #4 (#3)}

\def\etal{{\it et al.}}
\def\APH{\em Annals Phys.}

\def\CMP{\em Comm. Math. Phys.}

\def\DAM{\em Discrete Applied Math.}

\def\EPL{\em Europ. Phys. Lett.}
\def\FOP{\em Found. Phys.}

\def\GRG{\em Gen. Rel. Grav}

\def\IMD{{\em Int. J. Mod. Phys.} D}

\def\JHE{\em J. High Ener. Phys.}
\def\JMM{\em J. Math. \& Mech.}
\def\JMP{\em J. Math. Phys.}
\def\JPC{\em J. Phys. Conf. Series}

\def\JPS{\em J. Phys. Chem. Solids}

\def\MPA{{\em Mod. Phys.} A}

\def\NAP{\em Nature Phys.}

\def\NAT{\em Nature}

\def\NPB{{\em Nucl. Phys.} B}
\def\PHT{\em Physics Today}
\def\PHY{\em Physics}
\def\PLA{{\em Phys. Lett.}  A}

\def\PRA{{\em Phys. Rev.} A}

\def\PRD{{\em Phys. Rev.} D}
\def\PRL{\em Phys. Rev. Lett.}
\def\PRV{\em Phys. Rev.}
\def\PRE{\em Phys. Rep.}

\def\PTR{{\em Phil.Trans.Roy.Soc.Lond.} A}

\def\RMP{\em Rev. Mod. Phys.}

\def\SCI{\em Science}
\def\SHP{\em Stud.Hist.Philos.Mod.Phys.}

\def\SYN{\em Synthese}

\def\be{\begin{equation}}
\def\ee{\end{equation}}
\def\bea{\begin{eqnarray}}
\def\eea{\end{eqnarray}}
\begin{document}
\title{
{\begin{flushleft}
\vskip 0.45in
%{\normalsize\bfseries\textit{Chapter~1}}
\end{flushleft}
\vskip 0.45in
\bfseries\scshape Foundational role of symmetry in Quantum Mechanics and Quantum Gravity}}
\author{\bfseries\itshape Houri Ziaeepour$^{1,2}$\thanks{E-mail address: houriziaeepour@gmail.com}\\
$^1$ Institut UTINAM, CNRS UMR 6213, Observatoire de Besan\c{c}on, \\Universit\'e de Franche Compt\'e, \\41 bis ave. de l'Observatoire, BP 1615, 25010 Besan\c{c}on, France \\$^2$ Mullaerd Space Science Laboratory, University College London, Holmbury, St. Mary\\ Dorking GU5 6NT, Surrey, UK}

\date{}
\maketitle
\thispagestyle{empty}
\setcounter{page}{1}
% ------- [First Page Running Head] - place it immediately after title! ------
\thispagestyle{fancy}
\fancyhead{}
\fancyhead[L]{}%{In: Quantum Mechanics: Theory, Analysis, and Applications \\ 
%Editor:  } % needs \label{lastpage-01} on the last page.
\fancyhead[R]{}%{ISBN:  \\ \copyright~2017 Nova Science Publishers, Inc.}
\fancyfoot{}
\renewcommand{\headrulewidth}{0pt}
%------------------------------------------------------------------------------

\noindent \textbf{PACS:} 03.65.Ta, 03.65.-w, 04.60.-m, 05.30.-d, 05.70.Fh
\vspace{.08in}

\noindent \textbf{Keywords:} foundation of quantum mechanics; symmetry; quantum mechanics; quantum gravity; phase transition

%% Other situations:
\noindent \textbf{AMS Subject Classification:} 81B, 81C

\label{lastpage-01}

%\medskip
%\begin {tabular}{p{12cm}p{3cm}}
% & \bf \it {To Memory of Changoul and her Love}
%\end {tabular}
%\medskip

\begin {abstract}
Symmetry is a fundamental milestone of quantum physics and its applications such as quantum field 
theory of elementary particles and exotic states in condensed matter. However, symmetries are not 
explicitly involved in the definition of quantum mechanics axioms. Here we consider symmetry as a 
foundational concept and reformulate quantum mechanics and measurement axioms based on 
abstraction of physical entities by their symmetries. We argue that issues related to measurements 
and physical reality of states can be better understood in this description. In particular, the abstract 
concept of symmetry provides a basis-independent definition for observables. Moreover, we show 
that the apparent projection/collapse of quantum state as the final outcome of a measurement or 
decoherence is the result of breaking of symmetries. This phenomenon is comparable with a phase 
transition due to spontaneous symmetry breaking and makes the process of decoherence and 
classicality a natural fate of complex systems consisting of many interacting subsystems. Additionally, 
we demonstrate that the property of state space as a vector space representing symmetries is more 
fundamental than being an abstract Hilbert space, and its $L2$ integrability can be obtained from the 
imposed condition of being a representation of a symmetry group and general properties of 
probability distributions. This new description along with a recent proposal about the necessity of 
gravity for consistency and completeness of quantum mechanics may facilitate construction of a 
new model for quantum gravity, fundamentally different from previous candidates. As a first result of 
this approach we advocate an explanation for 3D perception of real space.
\end {abstract}

% ------------ [Running Heads - for odd and even pages] - please insert it only on page 2!
\pagestyle{fancy}
\fancyhead{}
\fancyhead[EC]{Houri Ziaeepour}
\fancyhead[EL,OR]{\thepage}
\fancyhead[OC]{Foundational role of symmetry in Quantum Mechanics and Quantum Gravity}
\fancyfoot{}
\renewcommand\headrulewidth{0.5pt} 
%------------------------------------------------------------------------------

\section {Introduction} \label{sec:intro}
The discovery and construction of a model for describing microphysics, namely the Quantum Mechanics, 
was a gradual process. It took over three decades for the model, which was first introduced as an 
empirical phenomenology to describe observations, to rise to status of an abstract model based on well 
defined axioms. Since that revolutionary epoch, which profoundly changed the very basis of our thinking 
about the physical world, every aspect of quantum mechanics and its predictions such as nonlocality, 
contextuality, superposition, and entanglement has been tested thousands of times, and up to precision 
of measurements, so far no inconsistency has been found. Nowadays quantum effects are not mere subjects 
of abstract interest for physicists, but the backbone of the $21^{st}$ century technologies: from 
electronic devices to drug design and quantum computers~\cite{qminfobook} However, despite being so far 
the most successful theory in the history of science, the principles of quantum mechanics are still considered 
not to be well understood. 

According to Wikipedia there are at least 14 well known, more or less distinct interpretations of 
quantum mechanics. There are also many more less-known interpretations, see 
e.g.~\cite{qmotherinterpret0,qmotherinterpret1,consistenthist} for some examples. 
Various Einstein-Podolsky-Rosen (EPR) type experiments have been performed to verify 
completeness of quantum mechanics and search for hypothetical hidden variables~\cite{hiddenvar} 
and test of other apparent paradoxes~\cite{hardyparadox0,hardyparadox1}. Moreover, a quantitative 
verification of quantum mechanics became possible after discovery of what is generally called 
{\it Bell's inequalities}~\cite{bellineqall}, after the original work by J.S. Bell~\cite{bellineq0,bellineq1}. 
Nowadays these experiments are carried out in different setups, including loophole free~\cite{qmnoloophole} 
and at long distances~\cite{eprlong0,eprlong1,eprlong2}, and use techniques which should lead to 
realization of quantum computing machines and quantum telecommunication. In addition, a recently 
proposed no-go theorem provides a mean to test hidden variables and incompleteness hypothesis 
with non-entangled particles~\cite{qmnooverlap} (from here on PBR). Furthermore, the test of 
contextuality using the Kochen-Specker theorem~\cite{qmcontextual}, which does not need 
specially prepared entangled particles, conclusively confirms predictions of quantum mechanics and 
their deviation from what is seen in classical systems~\cite{qmcontextualityex}.

The most remarkable and mysterious attribute of quantum mechanics is its nonlocality, which causes 
the violation of Bell's inequalities, the Hardy's paradox~\cite{hardyparadox0,hardyparadox1}, the 
PBR no-go theorem\footnote{This theorem assumes that copies of a system can be prepared 
independently. Properly speaking this violates nonlocality, but can be considered as a good 
approximation if correlation between systems is much smaller than measurement noise, which is taken 
into account in the construction and proof of the theorem~\cite{qmnooverlap}.}, and the inherent 
inconsistency of quantum mechanics with classical gravity~\cite{qmgrinconsist,houriqgr,bhfirewall}. 
Another controversial issue in quantum mechanics is the nature of a quantum state and whether it is 
a {\it real} physical entity or a representative of {\it the state of information} about a physical 
system available to an observer. Here {\it reality} means the complete representation, i.e. no hidden 
observable is averaged out. This implicitly means that a state should be {\it measurable} for 
each instance of a system. In classical physics, the state of a system is defined by a list of 
values $\{x\}$ measured or predicted for a minimal number of observables $\{X\}$ that characterize 
the system completely, that is the value of any other observable ${\mathcal O}$ accessible to an 
observer will be a single-valued function $o = {\mathcal O}(\{x\})$ such that:
\be 
o \neq o' \quad \Longrightarrow \quad [\{x\} = {\mathcal O}^{-1} (o)] \neq [\{x'\} = 
{\mathcal O}^{-1} (o')] \label{nooverlap}
\ee
It is important to emphasize on the single valuedness, because when multiple values are possible, 
some characterizing quantities should have been missed i.e. averaged out\footnote{In presence 
of chaos in a classical system the noise from environment plays the role of {\it hidden variable} 
and determines which branch is taken by the system, see e.g.~\cite{chaos}}. Classical statistical 
mechanics assumes that it is not possible to measure the complete set of characterizing observables 
$\{x_c\}$ of a system. The impact of averaging out unobservable (hidden) quantities is the random 
behaviour of accessible characteristics $\{x\}$ with a probability distribution $P(\{x\})$, and 
exclusion relation (\ref{nooverlap}) is not necessarily satisfied~\cite{bellineqall}. Thus, 
$P(\{x\})$ presents {\it the state of information} of the observer about the system. 

The no-go theorem of PBR proves that in quantum mechanics the state satisfies no-overlap 
condition (\ref{nooverlap}), and if quantum mechanics is a complete description of nature, the 
quantum state of a system is one of its physical properties. However, these authors conclude that 
the collapse/projection of state after a measurement is problematic for a physical property which 
should have {\it a reality of its own}~\cite{qmnooverlap}. This brings 
us to the well known measurement problem of quantum mechanics, namely the apparently nonunitary 
collapse of quantum states after a measurement~\cite{qmdecoherprob}. Although decoherence by an 
environment provides an alternative description for this enigma, 
see e.g.~\cite{decohererev0,decohererev1,decohererev2,decohererev3,qmdecoherbook,qftdecoher} for 
recent reviews and evidence for decoherence~\cite{qftdecoher} and ~\cite{qmmeasurrev} for review of 
dynamical approaches to measurement problem, it is not overwhelmingly accepted, and some questions and 
criticisms have been raised~\cite{decohcollapse,qmdecoherprob,qmphil}. For instance, it is 
argued~\cite{qmdecoherprob} that two possible outcomes cannot evolve from a single initial state 
from any {\it deterministic} unitary evolution. Therefore, stochasticity of possible outcomes 
cannot arise except in presence of hidden variables and correlations. However, this argument is 
based on the assumption of determinism, which is in contradiction with observations. Moreover, 
this reasoning does not apply to the case of a unitary but discontinuous evolution operator. 
Assuming a unitary evolution operator $U$ depending on a variable $v$ is applied to an initial 
state $|0\rangle$, if $U(v)$ is discontinuous at $v_0$, then: 
\be
|A\rangle = U(v_0 + \epsilon) |0\rangle \biggl |_{\epsilon \rightarrow 0} \neq 
|B\rangle = U(v_0 - \epsilon) |0\rangle \biggl |_{\epsilon \rightarrow 0} \label{unitarybifur}
\ee
A measurement apparatus can be defined as a system with such a property~\cite{qmsymmbreak}.

Criticisms against quantum mechanics~\cite{decohererev0,decohererev1,decohererev2,decohererev3,qmdecoherbook} can be summarized as the followings: 
\begin{description}
\item {\bf Meaning of a quantum state:} Similar to classical statistical physics, a quantum mechanical 
state defines a probability for measurement outcomes. But why does quantum mechanics seem to be a 
complete description of nature ?Why are Bell's inequalities violated ? Why, in contrast to classical 
statistical mechanics, does not the measurement of the probability of each outcome determine the state 
completely, decomposition coefficients are in general complex, create interference, and it is their 
norm square rather than themselves which correspond to probability of outcomes ? 
\item {\bf Measurement problems:} After a measurement, why do measured quantities behave 
classically, and immediate repetition of measurement gives the same outcome ? Collapse, entanglement, 
decoherence ? What is really measured and how is a pointer basis selected ?
\end{description}
The purpose of various interpretations of quantum mechanics is usually claimed to be clarifying 
these apparent issues, specially with the hope of constructing a theory conceptually closer to the 
macroscopic classical perception. 

In addition to numerous alternative interpretations, a number of authors 
have tried to construct quantum theory from a set of axioms that give it a structure very close to 
statistics. For instance, in a frequentist/measurement approach to classical statistical mechanics 
and quantum theory 5 axioms are proposed~\cite{qmhardy} from which 4 are satisfied by both theories 
and the last one only by quantum mechanics. Furthermore, in~\cite{qmqminfo} a similar construction 
inspired by quantum information theory with only 3 axioms is suggested. In both models the 
complex Hilbert space of quantum states is projected to a real vector space which presents 
probabilities for outcomes of measurements. Proposals of this type, collectively called 
{\it generalized statistical models}, try to interpret quantum mechanics as an extension of classical 
statistical mechanics. In~\cite{qmqminfo0} a model based on purely informational principles and 
operational probability is claimed to include quantum mechanics by adding one axiom to the 5 axioms 
of the model suggested in~\cite{qmhardy}. However, several issues such as the origin of 
probabilistic outcome of {\it circuits}, which are elementary objects of this model, is not 
discussed. Moreover, the state space of outcomes stays abstract and crucial issues such as observed 
absence of hidden variables and (non-)contextuality are not discussed. In fact their concept of 
coarse-graining should induce hidden variables and their first axiom {\it causality}, as it is 
described, seems to be in contradiction with the observed contextuality of quantum mechanics. 

Another approach for understanding measurement and decoherence issues is coupling of a classical 
system, e.g. a macroscopic measurement apparatus, with a quantum system. In this method either the 
classical system is extended to a quantum setting by applying superselection 
operators~\cite{qmclasssys} or the classical phase space is 
{\it complexified}~\cite{qmhamiltonext}. Then, the analogy between classical Poisson bracket and 
commutation relation is used to treat both systems in a similar manner. An application of this 
methodology~\cite{qmclassicalhybrid}, which satisfies consistency relations for performing the 
division to classical and quantum sub-systems of an ensemble is useful for studying large 
quantum systems. However, this approach fails its original aim of showing that quantum mechanics 
is some sort of generalization of classical mechanics. For instance, the characteristics of an 
ideal {\it deterministic} measurement apparatus obtained in~\cite{qmhamiltonext} is nothing else 
than what is called {\it quantum nondemolition measurement}~\cite{qmnondemol}. We should remind 
that numerous analogy between quantum and classical systems are found and can be used for  
study and simulation of quantum systems, see e.g.~\cite{qmclassicalanal}. An example is an 
invertible map between Hamiltonian of integer valued classical automata and bandwidth limited 
harmonic oscillator solutions of Schr\"odinger equation~\cite{automata}. However, similarities 
are limited to mathematical descriptions and no classical analogue for purely quantum effects 
such as entanglement and contextuality is known. 

Here we review a proposal for reformulating quantum mechanics around the fundamental concept of 
symmetry~\cite{houriqmsymm,houriqmsymm0}. We show that the issues and questions raised about quantum 
mechanics can be better clarified if we describe the relation between a system and an observer or 
environment by symmetries and the process of measurement\footnote{Here by measurement we always mean 
{\it sharp measurements} unless it is explicitly mentioned otherwise.} as their breaking. This 
interpretation does not modify the established principles of quantum mechanics and the process of 
measurement~\cite{qmvonneumann,qmdirac}. For this reason, we prefer to call it a new 
{\it description} or {\it language} for presenting quantum mechanics rather than a new 
interpretation or construction. 

The concept of symmetry is not new for quantum mechanics and quantum information. In algebraic 
quantum mechanics~\cite{qmvonneumann,qmalgebr} symmetries of state space (Hilbert space) 
and their relation with their classical analogues are extensively used to demonstrate the appearance 
of superselections by decoherence or lack of common reference frame~\cite{qmrefframe}. However, 
symmetries are not considered as a foundational necessity when issues related to measurement are 
addressed~\cite{qmmeasuresupersel,qmalgebrinterp}. In~\cite{qmsymminterp} the crucial role of 
symmetries in quantum mechanics is recognized, and superposition and indeterminism are associated to 
symmetries. But no axiomatic construction for quantum physics based on symmetries is proposed. 
Moreover, their emphasis is on the symmetries of spacetime coordinates, which are considered to have 
their own properties without any connection to quantum physics, and quantum effects are considered 
to be manifestations of relativistic spacetime symmetries. 

We present a systematic and axiomatic description of quantum mechanics in which symmetries and their 
representations have a central place in the construction of the theory and in the interpretation of 
physical world. In fact symmetries can be considered as extension of the fundamental logical concept 
of equality. They provide the necessary logical and mathematical tool for understanding both 
{\it static} concepts such as the nature of state space; and {\it dynamical} processes such as: 
evolution of state of a system, which satisfies Schr\"odinger equation (or Dirac/Klein-Gordon 
equation in relativistic cases); probabilistic outcome of measurements; interference; and nonlocal 
entanglement between components, which do not have a counterpart in classical statistical physics. 

Phase transition is proposed by many authors as a solution of measurement issues in quantum 
mechanics, see e.g.~\cite{qmsymmbreak,qmsymmbreak} and~\cite{qmmeasurrev} for a review. Nonetheless, 
symmetry description of quantum mechanics provides an ideal framework for this approach. Notably, we use a well 
known theorem by Sinai to prove that measurement and decoherence can be treated as phase transition 
following the breaking of a symmetry. We should remind that in contrast to description 
of~\cite{qmsymminterp}, spacetime symmetries do not present a special or privileged case. 
In particular, quantum systems can be separable in spacetime but entangled (inseparable) through 
another symmetry, for instance their spins. In fact, nonlocality of quantum mechanics is a natural 
consequence of allowed orthogonality of these symmetries. Furthermore, many of paradoxes and 
mysteries of quantum mechanics can be better understood in this new description. We will briefly 
explain a few of them in this review. We specially concentrate on topics which may have 
application in cosmology and quantum gravity. In particular, based on some preliminary ideas about 
what might be called a quantum vision of gravity~\cite{houriqgr}, we describe how the abstract 
concept of symmetry may help construct a quantum model which may simultaneously explain properties 
of spacetime and elementary particles.

In Sec. \ref{sec:qmaxioms} we reformulate axioms of quantum mechanics in symmetry language. 
Their content and consequences for properties of state space is described in Sec. \ref{sec:state}. 
The state space of composite systems and how it is related to state space of components, specially 
in what concerns their symmetries is reviewed in Sec. \ref{sec:composite}. In Sec. \ref{sec:measure} 
measurement and its relation with breaking of symmetries is discussed. In Sec. \ref{sec:random} 
we prove that probabilities for outcomes of measurements must satisfy the Born rule, and finalize 
the demonstration that the state space of a quantum system is indeed a Hilbert vector space. In the 
framework of standard quantum mechanics Hilbert space is usually a prior of the theory and although 
in practice it is always defined according to symmetries, it remains an abstract vector space in 
the axioms of the theory. On the other hand, in symmetry description, axioms of quantum mechanics 
associate it to symmetries and their realization by physical systems. In Sec. \ref{sec:decohere} we 
explain issues related to decoherence and classicality in the framework of symmetry description. 
Finally, in Sec. \ref{sec:qmgr} we briefly review arguments about inherent relation between quantum 
mechanics and gravity and present some ideas which may lead to understand the nature of spacetime 
and a consistent inclusion of gravity in quantum field theory. A short outline and prospective for 
future developments is presented in Sec. \ref{sec:summary}.

\section{Quantum mechanics postulates in symmetry language} \label{sec:qmaxioms}
In this section we present new description for axioms of quantum mechanics and compare them with 
their analogues \`a la Dirac~\cite{qmdirac} and von Neumann~\cite{qmvonneumann}, which from now on we 
call the {\it standard quantum mechanics}:
\renewcommand{\theenumi}{\roman{enumi}}
\begin{enumerate}
\item A quantum system is defined by its symmetries. Its state is a vector belonging to a projective 
vector space called {\it state space} representing its symmetry groups. Observables are associated 
to self-adjoint operators and a set of independent observables is isomorphic to a subspace of 
commuting elements of the space of self-adjoint (Hermitian) operators acting on the state space. 
%This subspace generates the maximal abelian subalgebra of the algebra associated to symmetry group of the system. 
\label{poststate}
\item The state space of a composite system is homomorphic to the direct product of state spaces of 
its components. In the special case of non-interacting (separable) components, this homomorphism 
becomes an isomorphism. The symmetry group of the composite system is a subgroup of direct product 
of its components. \label{postcomposite}
\item Evolution of a system is unitary and is ruled by conservation laws imposed by its symmetries and 
their representation by state space. \label{postunitary}
\item Decomposition coefficients of a state to eigen vectors\footnote{To be more precise we should 
use the term {\it rays} in place of vectors because vectors differing by a constant are equivalent. 
Thus, we assume that states are normalized.} of an observable presents the symmetry/degeneracy of 
the system with respect to its environment according to that observable. Its measurement is {\bf by 
definition} the operation of breaking this symmetry/degeneracy. The outcome of the measurement is 
the eigen value of the eigen state to which the symmetry is broken. This 
spontaneous\footnote{We explain in Sec. \ref{sec:symmbrdeco} why the breaking of degeneracy between 
states should be classified as spontaneous.} symmetry breaking reduces the state space (the 
representation) to subspace generated by other independent observables.\label{postmeasure}
\item A probability independent of measurement details is associated to eigen values of an 
observable as the outcome of a measurement. It presents the amount of symmetry/degeneracy 
of the state before its breaking by the measurement process. Physical processes which determine 
these probabilities are collectively called {\it preparation}\footnote{Literature on foundation of 
quantum mechanics consider an intermediate step called {\it transition} between preparation and measurement. 
Here we include this step to preparation or measurement operations and do not consider 
it as a separate physical operation.} . \label{postsymmbr}
\end{enumerate}
These axioms are very similar to their analogues in the standard quantum mechanics except that we 
do not assume an abstract Hilbert space\footnote{A Hilbert space $\mathcal{H}$ is a complete inner 
product vector space. Completeness is defined as the following: if the convergent point of all series 
of vectors $\sum_{k=0}^\infty u_k,~u_k \in \mathcal{H}$ which converge absolutely, that is 
$\sum_{k=0}^\infty\|u_k\| < \infty$, belongs to $\mathcal{H}$.}. We demonstrate this property of state spaces 
using axioms (\ref{poststate}) and (\ref{postsymmbr}) in Sec. \ref {sec:random}. In addition, we 
emphasize on the presence of symmetries in any quantum system and the fact that they distinguish 
one system from others. Furthermore, these postulates introduce a definition for measurement as 
operation of breaking a symmetry, partially or totally, and independent of how it is performed, that is 
by a designed apparatus or by interaction with another system or the rest of the Universe - the environment. 
It may look like a projection/collapse, i.e. a jump as it is the case in first-order phase transitions, or be 
continuous through entanglement with environment and decoherence. Additionally, we will show that 
the abstract concept of symmetry provides a basis-independent mean for describing what is observed.  
It can explain the problem of selected pointer basis during a measurement and ambiguity of 
observed quantities raised in the literature~\cite{qmvonneumann,qmalgebr}. 

In standard quantum mechanics, measurement axioms attribute a probability to 
projection/collapse/decoherence of a state $|\psi\rangle$ to a state $|\phi\rangle$ equal to 
$tr (\rho_\psi P)$ where $\rho_\psi$ is the density operator associated to the state $|\psi\rangle$ 
and $P_\phi$ is the projection operator to state $|\phi\rangle$. Here we will obtain this relation 
from properties of probability distribution functions and their application to composite systems 
described under axiom (\ref{postcomposite}).

\begin{figure}
\vspace{-1cm}
\begin{center}
\hspace{2cm}\includegraphics[width=15cm]{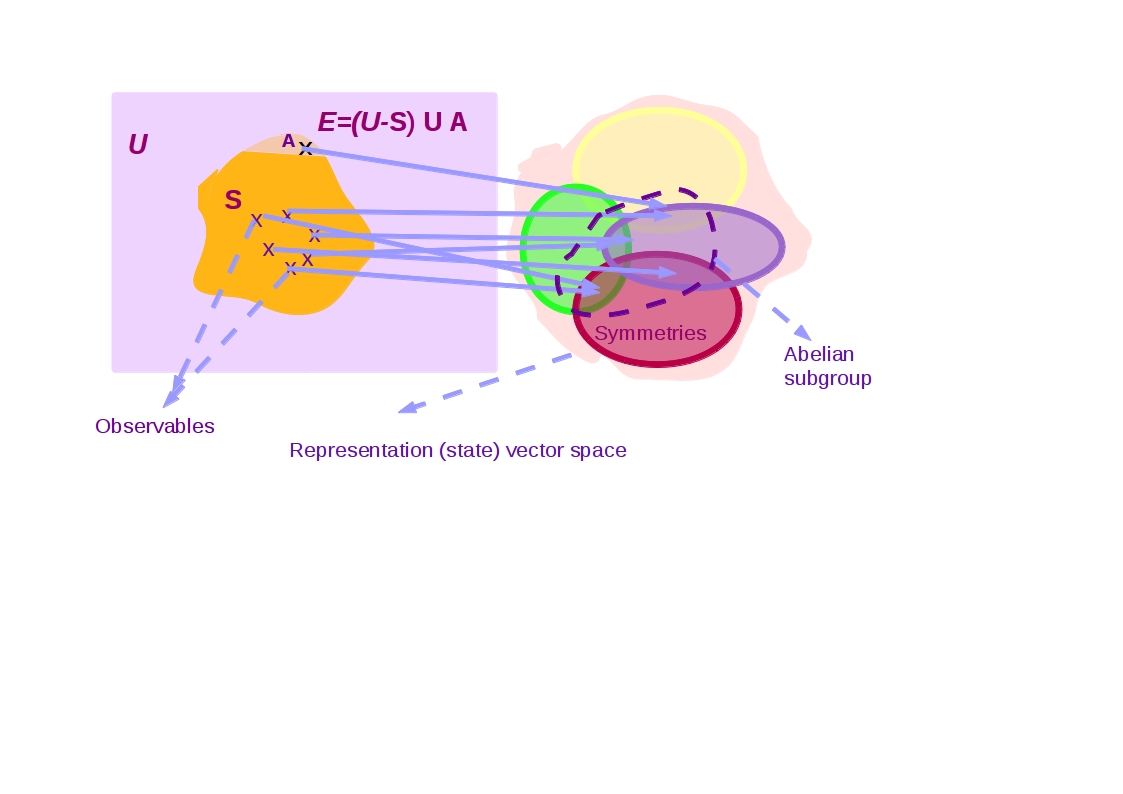} 
\vspace{-4cm}\caption{Abstraction of systems by their symmetries. The region $A$ indicates the 
shared/entangled information between system and environment (or observer) usually called apparatus. 
The overlap - interaction - between these subsystems is necessary, otherwise they cannot exchange 
information and each of them can be considered as a separate universe. Each observable / property 
(cross symbols) is associated to a symmetry group. The union of representation vector spaces for 
symmetries is called the {\it state space}. There is a one-to-one isomorphism between the maximal 
abelian subgroup and observables presented by projection arrows from system to representation 
(state) vector space \label{univdiag}}
\end{center}
\end{figure}

Symmetry can be considered as the generalization of equality and similarity concepts which are the 
basis of our logical and physical understanding of the Universe, see e.g.~\cite{mathlogicsymm} for 
symmetries in logic. The operation of measurement is nothing else than determining similarity between 
a characteristic of an object/system with one belonging to a reference. In fact, symmetry is our 
{\it only} tool for abstraction of the Universe. A question that arises here is whether it is 
necessary to assume a division of the Universe into {\it system} and {\it observer} or 
{\it environment}. In a logical view, the concept of similarity makes sense only if there are at 
least two objects to be compared. In a mathematical view and in the framework of set theory, which 
we assume to be applicable to all objects, including the Universe, only an empty set does not have 
subsets. In this sense, any indivisible set is either a subset of a larger set, or is isomorphic 
to an empty set, and thereby it is meaningless to discuss its properties. 
Consequently, a nontrivial Universe must have at least two non-empty nontrivial subsets, which can 
be called {\it system} and {\it observer} or {\it environment}. This argument shows that the 
necessity of having an observer/environment is not because without it quantum mechanics becomes 
meaningless\footnote{In explanations such as decoherence by a large environment and quantum 
Darwinism the large number of degrees of freedom 
in the environment has a central role for 
describing the removal of superposition by measurement or interaction with environment.} but 
because a subsystem\footnote{In this review a subsystem is defined as an entity in an irreducible 
representation of a symmetry group $G$. If it is not in a irreducible representation, it can be 
in turn decomposed to subsystems.} is required for assuring that Universe is not trivial. 
Fig \ref{univdiag} shows a schematic presentation of this description.
In the following sections we describe in more details axioms \ref{poststate}-\ref{postsymmbr} 
and how they help understand enigmatic issues summarized in the Introduction.

\section{State space and symmetries} \label{sec:state}
Symmetries are inherent to laws of physical. Examples are Poincar\'e symmetries of spacetime, gauge 
symmetries of fundamental interactions, global family symmetry of elementary particles, etc. 
They are intrinsic and preserved unless dynamically broken by specific interactions. 
Although in standard quantum mechanics the Hilbert space is an abstract object introduced by 
axioms, in practice it is always related to symmetries of the system and represent them. For 
this reason in axioms \ref{poststate}-\ref{postsymmbr} the emphasis is on symmetries. They 
provide a foundation for abstract definition of physical systems and their states. According to axiom 
(\ref{poststate}) a physical system - a subset of the universal set of the Universe - is abstracted 
by its symmetries and their representations, which define the vector space of states. Symmetries 
of physical systems are dominantly (if not always) Lie groups. Therefore, for the sake of simplicity 
through this review we restrict our arguments to them. This category also includes discrete crystal 
groups which are subgroup of continuous Lie groups.

Vector spaces representing symmetry groups are generated by eigen vectors of commuting 
subgroup, which according to axiom (\ref{poststate}) defines independent observables. Evidently the 
decomposition of an arbitrary state vector depends on the selected basis. Nonetheless, this does not 
affect the fact that another basis corresponds to eigen vectors of commuting elements of the 
symmetries too, and a change of basis only modifies the decomposition of states (vectors) not their 
identity or nature. Therefore, in this description observables have an identity independent of the 
chosen basis. For this reason, the question raised in Sec. \ref{sec:intro} about the pointer basis 
is irrelevant. For example, spin of particles are related to $SO(3) \cong SU(2)$ symmetry where
the abelian subalgebra of its associated algebra is 1-dimensional. Because $SO(3)$ is the rotation 
group in a 3D space, the generator of abelian subalgebra\footnote{Through this review 
{\it subalgebra of a group} should be understood as {\it subalgebra of the associated algebra 
of a group}.} is usually associated to one of the frame axes. Nonetheless, any other vector can 
be equally considered as the generator of the abelian subalgebra. What matters is the concept 
of commuting subgroup, not its basis-dependent representation.

\subsection {State symmetry} \label{sec:repgen}
Symmetry groups have many representations. Degeneracies of a system determines the representation 
realized by its state space. In turn, degeneracies can be considered as an {\bf induced symmetry} 
that we call {\it state symmetry}. For example, consider a $\pi^0$ particle decaying to two 
photons with opposite spins as system, and the rest of the Universe as environment/observer. 
In presence of a magnetic field in the environment the interaction of the field with charged quark-antiquark of $\pi^0$ defines a 
preferred direction for spins (helicities) of oppositely propagating photons. This breaks spherical 
symmetry of pion's environment and gives the observer some information about the spin (helicity) of 
e.g. the photon emitted to the hemisphere in the direction of magnetic field. If the field is very 
strong, the photon can have only one polarization. This means that its state space would be a trivial 
representation of spin $SU(2)$ symmetry. In absence of an external field no preferred direction 
is present and the spin of each photon at production is completely arbitrary. In this case the state 
has a spherical symmetry reflected in the equal coefficient of components when the state is 
decomposed to eigen states of the maximal abelian subgroup of $SU(2)$. Furthermore, this 
symmetry/degeneracy of spin states is independent of other properties, notably geometry and distance 
between particles, and is a global attribute. 

Cautious is in order when identifying symmetries of a system and their representations. The simplest 
way of discussing this issue is through an example: Consider the electron in a hydrogen atom. The 
nucleus (proton) creates a spherically symmetric electric potential in which the single electron of hydrogen  
is moving. Then, what is the symmetry of electron as a system when the aim is measuring its position ? 
A naive view may conclude a spherical $SO(3)$ symmetry. Indeed, this is the symmetry of Lagrangian or 
Schr\"odinger equation for electron. On the other hand, in the electron frame the electric potential is not 
spherically symmetric. Only if the electron can {\it see} the entire space, that is when it is not 
localized, the potential and space look spherically symmetric. The only symmetry - degeneracy 
or degrees of freedom - related to the position of electron is translation symmetry 
in $\mathbb {R}^3$ space. Evidently this is not a perfect symmetry (degeneracy) because the 
presence of a positive electric charge prefers points closer to the nucleus rather than further. 
Therefore, the state of the electron is represented by the infinite-dimensional abelian 
translation symmetry $\mathbb {T} \times \mathbb {T} \times \mathbb {T}$ where $\mathbb {T}$ is 
1D translation group. Due to rotational symmetry of potential eigen states of position operator - 
obtained from solution of Scht\"odinger equation - have simple description when they are 
decomposed to radial and angular components, and solutions are classified by two $\ell$ and $m$ 
integers. Nonetheless, a probability must be associated to position of electron, e.g. in spherical 
reference frame $(r, \phi, \theta) \in \mathbb{T} \times \mathbb{T} \times \mathbb{T} \cong U(1) 
\times U(1) \times U(1)$. This example shows that the state vector space represents symmetries 
which define the degrees of freedom of a system rather than its dynamics which is related to its 
interactions. In standard quantum mechanics degeneracies presented  by the state space are treated as lack of 
information or in another word entropy, which is mathematically related to the density matrix. On
the other hand, viewing these degeneracies as symmetries opens up the possibility of new 
abstraction and interpretation of quantum puzzles.

%\subsection {Properties of state space} \label{sec:repprop}
Once an observable is {\it measured} (we define this operation in Sec. \ref{sec:measure}), 
according to axiom (\ref {postmeasure}) the state symmetry or degeneracy between eigen vectors 
breaks, and thereby the representation is reduced to trivial. State symmetries (degeneracies) are 
related to system's history, initial conditions and/or preparation. Consequently, they are 
ephemeral and may be total or partial. In the former case coefficients of decomposition of the 
state to eigen vectors are equal. Otherwise, the symmetry is partial. The system's 
history or environment somehow discriminates between eigen states, but cannot single out one of 
them. The symmetry of states is by definition associated to and inseparable from 
system-environment symmetries. Therefore, breaking of degeneracies is comparable and analogous to 
spontaneous symmetry breaking, in which the symmetry is broken in states but not (or softly) in the 
Lagrangian (see Sec. \ref{sec:measure} for more details).

%According to axiom (\ref{poststate}), state space is a vector space representing the symmetry group of the system, and observables are self-adjoint linear operators acting on this space. By definition for every group element $g \in G$ there is an inverse element $g^{-1} \in G~|~gg^{-1} = g^{-1}g = I$. 
Only in vector spaces defined over complex numbers $\mathbb{C}$ operators with non-zero determinant 
are always diagonlizable\footnote{Strictly speaking this is true for finite representations.}. This 
property is necessary to ensure the existence of a basis composed of eigen vectors for every 
observable. Because properties of a system are by definition conserved under application of its 
symmetry group $G$ and observables are self-adjoint (hermitian) operators, the representation of 
symmetry groups of the system by its state space must be unitary. This requirement reinforces the 
necessity of the definition of state space over $\mathbb{C}$ field. Although 
many groups are unitarily represented by vector spaces defined over real 
numbers, complete set of representations are realized on a complex field.\footnote{Representations 
over $\mathbb{C}$ are isomorphic to real representations with double dimensions, which are not minimal 
and can be considered as alternative definition of a complex representation. They are analogous to 
presenting complex numbers by their real and imaginary components.}. In standard quantum mechanics 
similar procedures are used to define a Hilbert space, but the role of symmetries is not considered 
as a foundational aspect.

\subsection {Active manifestation of symmetries in quantum mechanics} \label{sec:eigen}
Application of a group member transfers a state:
% If $U_1 \& U_2 \in M_A$ where $M_A$ is the maximum abelian subgroup, eigen vectors of U_1 are also  eigen vector of U_2: $U_1 |\alpha\rangle = \alpha |\alpha\rangle \Longrightarrow U_2U_1 |\alpha\rangle = U_1U_2 |\alpha\rangle = \alpha U_2|\alpha\rangle$. Thus $U_2|\alpha\rangle$ is also eigen vector of $U_1$. Thus in general - unless there are degenarate eigen values - $U_2|\alpha\rangle \propto |\alpha\rangle$, i.e. $U_2|\alpha\rangle = cte. |\alpha\rangle$.
\be
|\psi\rangle = \sum_\alpha a_\alpha |\alpha\rangle, \quad |\alpha\rangle \rightarrow 
|\alpha'\rangle = U |\alpha \rangle,~ U \in G \quad \Longrightarrow \quad |\psi\rangle \rightarrow 
|\psi'\rangle = \sum_{\alpha'} a_\alpha |\alpha'\rangle \label{decomptrans}
\ee
Here $G$ is the symmetry group of the system and we assume that $|\alpha\rangle$ is a complete 
set of eigen vectors of the maximal abelian subspace $G_A \subset G$. Application of a symmetry 
transformation projects $G_A$ to a new subspace but preserves the commutation relation between 
its members. The new states $|\alpha'\rangle$ are now eigen states of $G_A$ . This means that 
$|\alpha\angle$ states are no longer eigen vectors of the new abelian subspace. Therefore, 
coefficients $a_{\alpha}$ do not have the same physical interpretation with respect to the new 
representation of the subgroup $G_A$.

To better understand the physical process described by (\ref{decomptrans}), consider a successive 
Stern-Gerlach experiment. Electrons are polarized by passing through the first Stern-Gerlach 
detector. Then, only one of two polarizations, e.g. $Z^+$ is kept and we identify their state with 
$|\psi\rangle$ in (\ref{decomptrans}). The set of eigen states $\{|\alpha\rangle\} \equiv 
\{|Z^+\rangle, |Z^-\rangle\}$ correspond to two polarizations along $Z$ axis. After passage of 
electron through the first detector $a_{Z^+} = 1,~a_{Z^-} = 0$. The passage of these 
electrons through the second detector in which the magnetic field is in e.g. $X$ direction is equivalent 
to application of a symmetry transformation, in this case the application of $\hat{S}_x$ or its 
representation as Pauli matrix $\sigma_x$ on the state $|\psi\rangle$. In fact the presence of 
a magnetic field in $X$ direction rotates the preferred direction for electrons, which before 
entering to the second detector was $Z^+$, to $X^+$ direction. As explained above this operation 
projects the abelian subspace to the subspace in the $X$ direction, and according to postulate 
(\ref{poststate}) observables become members of this new abelian subspace generated by eigen 
vectors of $\hat{S}_x$. In agreement with experiments, a simple calculation shows that the expansion 
of $|\psi'\rangle$ with respect to eigen vectors of $\hat{S}_x$ will include both possible polarizations 
with equal probabilities when electrons leave the second detector. The transformation of 
state space by a member of symmetry group can be also seen as the absence of a common basis 
for the definition of the state space by two observers, see e.g.~\cite{qmrefframe}.

Another set of noncommuting measurables, which have wondered physicists, are position and momentum. 
They are both related to translation symmetry, but it seems that the collapse/projection of wave function 
of macroscopic objects prefers the position basis. This even has led to interpretation of quantum 
mechanics / quantum gravity with position as preferred observable~\cite{causalset}. On the other 
hand, momentum basis is considered to be the preferred description in another regime of quantum 
gravity~\cite{qgrmom,qgrmom0}. Considering a fundamental preference for one or the other 
presentation is evidently a misinterpretation. What we actually measure is 
$\hat{X}(t) - \hat{X}(t_0)$ which commutes with $\hat{\vec{P}}(t) - \hat{\vec{P}}(t_0)$ (here we 
have used Heisenberg picture). Most objects are non-relativistic. Thus, 
$\langle \hat{X}(t) - \hat{X}(t_0) \rangle \approx 0$, i.e. they look to be in a position eigen 
state. Moreover, the concept of a limit velocity is a classical view and does not apply to off-shell 
relativistic particles. Therefore, the rest frame of relativistic particles is not special.

In conclusion consideration of active role of symmetries solves what is called {\it preferred basis 
issue}. In contrast to classical mechanics, in quantum systems symmetries have an operative role 
and induce degeneracies - through non-trivial representation of symmetry groups by the state 
space - and reflects history and relation of a system with its environment.

\subsection {Comparison with classical statistical mechanics} \label{sec:compclass}
In classical statistical mechanics a system with $n$ observables has a configuration 
space isomorphic to $\bigotimes^n U(1) \cong \mathbb R^n$ irrespective of symmetries 
and intrinsic symmetries - conservation relations - define a set of equivalence classes. 
In this space, a state is a point rather than a vector, and probabilities 
are $n-1$ simplexes in $\mathbb R^n$~\cite{qmdecoherbook}. Symmetries have passive roles. 
They reduce the number of degrees of freedom and make distribution of some observables similar. 
For instance, spherical symmetry means that properties such as distribution of particles and their 
density is spherically uniform. But this property does not induce any constraints on observables. 
In particular, states do not present a representation of $SO(3)$ symmetry group, which needs a 
vector space.

It is noteworthy to remind that in general, symmetry groups are non-abelian and their maximal 
abelian subalgebra presenting independent observables of quantum systems has a smaller dimension 
than the whole group. Therefore, quantum systems have less independent observables than their 
classical analogues. This is another manifestation of the active role of symmetries in quantum 
mechanics. They impose constraints on quantum systems absent in their classical counterpart.
Considering as an example the case of rotation symmetry discussed in the framework of Stern-Gerlach 
experiment, in classical statistical mechanics, if the direction of the angular momentum cannot 
be known from preparation or initial conditions, the system has a rotational $SO(3) \cong SU(2)$ 
symmetry, which is reflected in the equal probability for the projection of its angular momentum 
on the three reference directions. In analogy with quantum mechanics we can define a 
configuration (state) space consisting of states $|S_x,~S_y,S_z \rangle_c$, where subscript $c$ 
means classical. The state of the system can be described as:
\be
|\psi \rangle_c = \sum_i f_i |i\rangle_c \quad \quad i = \{S_x,~S_y,S_z\} \label{classspin}
\ee
where $S_x,~S_y,~S_z$ are components of angular momentum necessary for uniquely determining 
classical states. One can even {\it quantize (discretize)} outcomes by considering a 
{\it resolution} $\Delta S$ for projections. It presents the smallest detectable difference between 
two states. We define coefficients $f_i \geqslant 0$ as the probability for state $i$. The 
spherical symmetry is reflected in the invariance of $f_i$'s under permutation of axes and a 
rotation of the reference frame which projects $f_i$'s to themselves. The symmetry represented by 
l$|\psi \rangle_c$ states is $\mathbb{T} \times \mathbb{T} \times \mathbb{T} \cong U(1) 
\times U(1) \times U(1)$ if the amplitude of angular momentum is unknown, otherwise it is 
$\mathbb{T} \times \mathbb{T} \cong U(1) \times U(1)$. The spherical symmetry is present, but 
plays a passive role, namely it is reflected in the similarity of probability distribution of 
components, but it does not induce any correlation between their values.

\section{Composite systems, interactions, and environment} \label{sec:composite}
According to axiom (\ref{postcomposite}) the ensemble of two or more subsystems can be considered 
as a system with a symmetry group homomorphic to direct product of symmetries of its components. 
Deviation of this homomorphy from an isomorphy presents the strength of interaction or quantum 
correlation between subsystems. We remind that quantum correlation or in another word 
{\it entanglement} needs direct or indirect interaction. In the latter case the classical 
Lagrangian of the system may include no interaction between some components, but they can become 
entangled through quantum correction from their interaction with components which are in 
interaction with each others. Another possibility is an interaction in the past, that is their preparation. 
Thus, the axiom about composite systems is particularly important for understanding entanglement, 
measurement problems, and decoherence of quantum systems. In this regard, symmetries provide 
a natural criteria for division of the Universe to subsystems/components according to their symmetries. 

In the example of a decaying $\pi^0$ in Sec. \ref{sec:repgen}, the system is composed of three 
subsystems each with $SU(2) \cong SO(3)$ symmetry. Their ensemble has $SU(2) \oplus 
\mathbb{Z}_2 \subset SU(2) \times SU(2)$ symmetry where the first $SU(2)$ corresponds to 
the global symmetry under rotation and the second one the spin symmetry of photons. 
The division of this system to its components in position space is valid only when photons are 
sufficiently far, that is at perturbative distant scales from their production place. Nonetheless, 
as it is well known due to $\mathbb{Z}_2$ symmetry photons remains correlated - entangled - 
at any distance without any causal interaction between them, because their spin $\mathbb{Z}_2$ 
symmetry is orthogonal to coordinates shift symmetry, and induces what Einstein famously called: 
{\bf Spooky interaction at distance}.

In standard quantum mechanics mixed states are defined by their non-idempotent density matrix: $\rho^2 \neq \rho$. 
In symmetry description the omnipresence of symmetry allows to define them in a manner which 
highlights their incompleteness. The homomorphism, which according to postulate \ref{postcomposite} 
present a composite system reduces the symmetry to $G \subset G_1 \otimes G_2$. A density 
matrix for a pure state belongs to the space of operators $\rho_{pure} \in \mathcal{B}(G)$ where 
$G$ here must be understood as vector space representing the symmetry group of the composite 
system. Mixed states are defined as $\rho_{mixed} \in V \subset \mathcal{B}(G)$. Therefore: 
$T_g \rho_{mixed} T^\dagger_g \notin \mathcal{B}(G)$. It is straightforward to prove that this  
definition satisfies non-idempotent condition. Moreover, this definition explicitly demonstrates the 
incompleteness of information about the composite system. For instance, partial trace over 
density matrix of a pair of particles entangled by their spins is non-unitary - does not preserve 
probabilities - and restricts $Z_2$ symmetry of projection $Z_2 \times \mathbb{Z}_2 \rightarrow 
\mathbb{Z}_2$ to the subspace generated by the remaining particle.

\subsection{Inseparability of the Universe} \label{sec:insepar}
From close relation of fundamental interactions and symmetry groups, and from universality of 
gravitational interaction - for which we do not yet have a quantum description - we conclude that 
the decomposition of the Universe to subsystems is not orthogonal and it is not possible to 
decompose the Universe to two quantum subsystems $S_1$ and $S_2$ such that 
$S_1 \cap S_2 = \oslash$ and $\mathcal{H}_{S_1\cup S_2} = \mathcal{H}_{S_1} \otimes \mathcal{H}_{S_2}$. 
Therefore, quantum correlation is omnipresent, and for this reason the 
state space of the Universe consists of an ensemble of topologically open and algebraically 
intertwined vector spaces. This means that for any system $S \subset U$ there is a system $S'$ 
such that $S \subset S' \subset U$. But, if $S' = S \cup S_1$, its state space 
$\mathcal{H}_{S'} \neq \mathcal{H}_S \otimes \mathcal{H}_{S_1}$ and quantum correlation between 
its subsets are never null. Therefore, division to system/apparatus/environment is always an 
approximation and the Universe is inherently nonlocal, composite but inseparable. This observation 
answers criticisms raised against decoherence as the origin of classical behaviour of macroscopic 
systems. Their claim is that decoherence only approximately suppress interference~\cite{decohcollapse}. 
However, arguments presented here show that classicality and locality of macroscopic world 
which are indeed approximations and consequence of a coarse-grained presentation of an 
otherwise quantic Universe. 

Because classicality is just an approximation, the Universe as a whole must be considered as a 
quantum system, and there should exist a state space representing its global symmetries. They must 
be related to symmetries of its components by a homomorphism. If the projection is trivial, the 
Universe will have a single state meaning that it globally looks classical. If it is roughly 
isomorphic, its state is simply the direct multiplication of state spaces of its components. If it 
is neither trivial nor isomorphic, but isomorphic to a small nontrivial symmetry, the Universe 
may have global structures. Observations of Cosmic Microwave Background 
(CMB)~\cite{cmbplanck0,cmbplanck1} do not show any signature of a nontrivial topology up to their 
precision and anisotropies seem to have a continuous spectrum up to modes $k \rightarrow 0$. This is 
topologically consistent with an open ball. Nonetheless, the Universe may be in a nontrivial 
rotational ($SO(3) \cong SU(2)$) state~\cite{cmbplanckaniso}.

\section{Measurement} \label{sec:measure}
A measurement operation, as perceived classically, consists of using a tool to determine 
the value of a quantity or occurrence of a configuration, such as head or tail as the state 
of a thrown coin. By definition such operations lead to one value for n observable or one 
state for the system. Due to causality, if all dynamical variables are known, evolution of 
classical observables and outcomes of measurements are fully predictable. For this reason 
randomness in statistical mechanics is assumed to be due to lack of information and 
averaging over inaccessible degrees of freedom.
 
Quantum systems share the concept of randomness with statistical mechanics but their 
properties, as discussed in the previous sections, are quite different. Notably, state of a 
quantum system is a vector, which in general cannot be directly measured, because 
measurements make the state collapse/decohere to an eigen (pointer) state. Moreover, 
it seems that it is not always possible to associate the outcome to a unique observable 
and measurements are {\it contextual}. Another apparent problem is the absence of 
superposition of pointer states as outcome of sharp measurements. In addition, 
repetition of a non-demolishing measurement reproduces the same outcome, 
see e.g.~\cite{decohererev2,decohererev3,qmphil} for examples of recent review of criticisms.

 In this section we explain the meaning of axiom (\ref{postmeasure}) that associates a 
measurement to breaking of symmetries, and explain how it may clarify measurement issues.

\subsection {Breaking of degeneracies} \label{sec:symmbreak}
According to postulate (\ref{postmeasure}) a  sharp measurement breaks symmetry and 
degeneracy of states, and reduces the state space to a subspace in which the operator 
presenting the measured quantity is represented trivially. In this sections we show that 
this process is similar to spontaneous symmetry breaking in macroscopic systems. 
Moreover, because independent observables are abelian, the reduction of algebra for one 
observable does not affect states of other independent observables. In fact, the algebra 
generated by independent observables can be divided to quasi-orthogonal subalgebras. 
Notably, the measurement of an observable which commutes with spacetime operators 
do not depend on the position state and is global. This explains the apparent paradox 
of EPR-type experiments. According to  postulate (\ref{postmeasure}) in EPR experiments 
the measurement of the spin of one of the entangled particles breaks spin $SU(2)$ symmetry, 
defines a pointer/preferred direction, and spin state becomes trivial, is no longer a 
superposition of eigen states. Then, the entanglement - quantum correlation - 
ensures that the state of the other particle is automatically fixed, irrespective of the distance between 
two particles, because $SU(2)$ spin symmetry and spacetime translation symmetry commute with 
each other.

\subsubsection{Apparent ambiguity of measured observables} \label{sec:obser}
In quantum measurements (and classical measurements when direct comparison with a reference is not 
possible) an interaction between system and measurement apparatus is necessary. It may explicitly 
breaks some of symmetries of the system. For instance, for measuring spin of a particle, interaction 
with an external magnetic field or spin of another particle is necessary. Assuming that the 
direction of the spin of the particle before measurement is unknown, its preparation had to be under 
a spherically symmetric condition. The interaction with a field inside the apparatus defines a 
preferred direction and explicitly breaks the spherical symmetry in system-apparatus Lagrangian. On 
the other hand, interaction with spin may or may not induce a preferred direction. For instance, 
if the interaction between two spins is antisymmetric, that is proportional to their vector product 
$\mathbf {S}_1 \times \mathbf {S}_2$ (Dzyaloshinskii-Moriya interaction~\cite{dzmspinint,dzmspinint0}), 
where indices $1$ and $2$ indicate system and apparatus respectively, the transverse plane to 
$\mathbf {S}_2$ is the preferred plane for measurement of spin. However, when the direction 
of the spin to be measured is known to be in the preferred plane, for example if the system consist of 
photons propagating in vacuum parallel to $\mathbf {S}_2$, there is no preferred direction on 
the plane for their spin. In the standard description of quantum mechanics the measurement of 
the spin of such photons by this measurement apparatus seems problematic and it is not 
clear which component of spin vector is measured. However, the outcome is always $1$ or $-1$ 
and not a value in between, as expected in a classical measurement. On the other hand, in symmetry 
description there is no ambiguity about what is measured: The observed quantity is the unique 
generator of abelian subalgebra of $SU(2)$ which is independent of arbitrary definition of 
coordinates on the plane. 

\subsection{Influence of environment and decoherence in symmetry description} \label{sec:measuredecoher}
Modern literature on the foundation of quantum mechanics present measurement as entanglement between 
system and apparatus such that the quantum nature of both be explicit~\cite{decohererev2,decohererev3}:
\be
\sum_\alpha s_\alpha |\alpha \rangle \otimes |a_0 \rangle \rightarrow 
\sum_\alpha s_\alpha |\alpha, a_\alpha \rangle \label{measentangle}
\ee
where $|a_0 \rangle$ is the initial state of the apparatus and $|\alpha \rangle$ is an arbitrary 
basis for the system. The interaction between two spins, one considered as system and other as 
apparatus, is an example for (\ref{measentangle}). 

In the framework of symmetry description, system and apparatus must have an interaction related 
to the symmetry represented by states $|\alpha \rangle$. For the sake of simplicity we assume 
that this representation is finite with dimension $N < \infty$. To distinguish between these 
states, the initial apparatus representation must be $M \geq N$ dimensional. The entanglement by 
interaction breaks the $[N] \otimes [M]$ dimensional representation of system-apparatus to an 
$[N]$ dimensional representation of the symmetry\footnote{More generally, the projection can be 
$[N] \otimes [M] \rightarrow [N']$. If $N' > N$, some outcomes would be equivalent. If $N' < N$ 
the apparatus will not be able to single out some states and they may create an interference.}. 
This process of entanglement and symmetry breaking is explicit and related to the design of 
the apparatus. However, some criticisms have been raised against this approach to 
measurement~\cite{decohererev2}. Their concern is that there is no guarantee 
$|\alpha, a_\alpha \rangle$ states be orthogonal to each others. As we discussed in the previous 
subsection, this cannot be the case because the right hand side of (\ref{measentangle}) 
always presents a member(s) of maximal abelian subspace associated to the symmetry in 
the representation realized by the subspace generated by 
$|\alpha, a_\alpha \rangle | \forall \alpha$. And by definition if the apparatus is able to perform 
a sharp measurement, the final state is one of orthogonal eigen states.

\subsection{Symmetry breaking and decoherence} \label{sec:symmbrdeco}
Entanglement approach does not clarify the final breaking of degeneracies in quantum systems 
after a measurement.  A possible solution for this issue could be the repetition of entanglement 
process by enlarging the apparatus in successive steps until the ensemble become 
a macroscopic classical apparatus/environment: 
\be
\sum_\alpha s_\alpha |\alpha \rangle \otimes |a_0 \rangle \otimes |e_0 \rangle \rightarrow 
\sum_\alpha s_\alpha |\alpha, a_\alpha \rangle \otimes |e_0 \rangle \rightarrow 
\sum_\alpha s_\alpha |\alpha, a_\alpha, e_\alpha \rangle \label{decoherentangle}
\ee
where $|e_0 \rangle$ is the state of a macroscopic environment. Because the state of environment 
is not accessible, it must be traced out. This operation makes the state mixed state with a 
non-idempotent density matrix $\rho \neq \rho^2$. If off-diagonal elements of density matrix are 
negligible with respect to diagonal elements, the ensemble system-environment behaves similar to 
a classical random system. Indeed, in the framework of algebraic approach to quantum mechanics 
it is demonstrated that when an infinite or very large number of degrees of freedom are traced out, 
a {\it superselection}~\cite{qmalgebr,qmmeasuresupersel,qmdecoherbook} occurs and the density 
matrix of the combined system-apparatus (or more generally two systems) becomes 
approximately diagonal in a basis which is chosen by the interaction between apparatus and 
environment (system-environment interactions are usually neglected, see also Sec. \ref{sec:decohere} 
for more details). 

\subsubsection{Decoherence by lack of common reference frame} \label{sec:refframe}
Superselection and decoherence occurs also in absence of a common reference 
frame or basis between 2 or more parties - here the system, apparatus/observer, and 
environment. It is proved, see~\cite{qmrefframe} and references therein, 
that in absence of a common reference frame a completely Positive Operator Valued 
Measurement $\mathcal{E}$ on a system prepared with density matrix $\rho$ by one of the parties 
is decomposed and decohered as the following for other parties:
\be
\mathcal{G} = \sum_q (D_{M_q} \otimes I_{N_q}) \circ P_q    \label{refdecomp}
\ee
where $q$ is the conserved charge of subsystems presenting irreducibly a symmetry group $G$: 
$\mathcal{H} = \bigoplus_q \mathcal{H}_q = \bigoplus_q M_q \otimes N_q$. In this decomposition 
$M_q$ and $N_q$ are virtual decomposition of state space to symmetry invariant and multiplicity, 
respectively. Operator $P_q$ projects $\rho$ to constant charge subspace; operator $D_{M_q}$ when applied to 
$M_q$ behaves as identity operator proportional to a constant; and finally, operator $I$ is identity and 
acts on the multiplicity\footnote{To be more precise, $P$, $D$, and $I$ are super-operators, 
that is they are operators which act on the space of operators acting on the state space noted as 
$\mathcal{B}(\mathcal{H}$.}. In \ref{refdecomp}) there is no quantum correlation 
between subspaces with different charge and they are decohered from each others. 
Thus, $\mathcal{G}$ behaves similar to a classical statistical ensemble. 
However, the absence of common reference does not affect subspaces invariant under 
$G$ and quantum correlation of these states are preserved. For this reason 
they are called {\it decoherence-free}~\cite{decoherfree} subspaces. We should emphasize that the 
presence of a symmetry and treatment of the space vector as its representation is crucial 
for proving \ref{refdecomp}). This is one more evidence that symmetries are inherent concepts 
in quantum mechanics. 

The relation between system and apparatus/environment can be seen as absence 
of a common reference frame and the above theorem applies and proves 
that the apparatus/environment sees the state of the system as a statistical ensemble.
However, these arguments do not resolve  the conceptual puzzle that in a sharp measurement one 
eigen state is selected among many others. This issue has raised criticisms against 
decoherence as a replacement for nonunitary collapse to a pointer 
state~\cite{decohcollapse,qmdecoherprob,decohererev2}. Nonetheless, decoherence seems 
to be the most physically motivated explanation for random selection of one eigen state by a 
measurement. Moreover, it is consistent with modern quantum experiments in which even 
macroscopic systems can be kept in entangled or superposition states for significant amount 
of time, if their interaction with environment can be prevented~\cite{macroentagle}.

\subsubsection{Decoherence by large environment}   \label{sec:largeenv}
In the framework of symmetry description, the large number of degrees of freedom of environment 
means that it realizes an infinite dimensional representation of the symmetry, to which the 
observable belongs\footnote{Unless the environment is in a very special quantum state, e.g. a 
condensate. We neglect exceptional cases here, but they may be important in cases such as 
decoherence in the early Universe.}. Assuming system\footnote{For sake of simplicity we include 
apparatus in the environment.} in an $N$-dimensional representation of the symmetry, according to 
postulate (\ref{postcomposite}) the composite system-environment is in $N \times \infty$ 
representation if there is no interaction/correlation between them, otherwise in a subspace of this 
representation. In particular, an entanglement reduces the state to:
\be
|\psi, E\rangle = \sum_{\alpha,~\epsilon^{\alpha}_i, i=0}^{i \rightarrow \infty} 
{f}(\alpha ,\{\epsilon^{\alpha}_i\}) | \alpha, \epsilon^{\alpha}_1, \epsilon^{\alpha}_2, \cdots, 
\epsilon^{\alpha}_i, \cdots\rangle \label{sysenventang}
\ee
where $\psi$ and $E$ indicate system and environment, respectively, and $\alpha$ and 
$\epsilon^{\alpha}_i, i=0,\cdots$ represent eigen vectors of the system and components of 
the environment, respectively. It is conceivable that there are large number of equivalent 
realizations of the symmetry in (\ref{sysenventang}) - corresponding to virtual multiplicity 
subspace $N_q$ in (\ref{refdecomp}).  For instance, the state $\alpha$ most probably 
entangles only to few spatially closest components of environment and others play the role of 
spectators. Because the degrees of freedom or components of the environment are not observed, states in 
(\ref{sysenventang}) form infinite equivalence classes of the representation of the symmetry realized 
by $\alpha$. They play a role similar to statistical ensemble in classical statistical mechanics, and 
induce a probability for occurrence of a specific $\alpha$ as outcome of a measurement. However,  
there is a significant difference with classical case. Random selection of one of the members of 
these equivalence classes breaks the symmetry and its realization by the system becomes trivial. 
Therefore, in contrast to classical case, immediate repetition of the measurement on the same system 
gives the same value for the observable. 

Although the environment plays the role of a hidden variable, equivalence classes are tagged by the 
system, and once the symmetry is broken, both system and environment are constrained to the 
selected value of the observable, and symmetry representation becomes trivial. Thus, symmetry 
and its breaking have crucial role in the existence of equivalence classes, decoherence, and 
repetition of outcome in successive similar measurements. Moreover, because equivalence 
classes realize the same representation of symmetry as the system, probability is determined 
by the symmetry/degeneracy of the state of system and is completely independent of the state of 
environment. 

\subsection{Measurement as a phase transition} \label{sec:symmbrspontan}
In quantum systems selection of pointer basis by symmetry breaking is causal. This fact has been 
experimentally proved by employment of analyzers in which the direction of measurement changes 
with time\cite{eprtimevar}. They show that the contextuality of quantum measurements is a direct 
consequence of active realization of symmetries by quantum systems and their breaking by interaction 
with environment. Moreover, in symmetry description the selection between eigen states is similar to 
spontaneous symmetry breaking because it is imposed to the system by constraints originated from 
its interaction with environment. Equation (\ref{decoherentangle}) and description of symmetry breaking 
by constraining effects of many degrees of freedom in Sec. \ref{sec:largeenv} show that the final 
decoherence of quantum systems during a measurement is independent of details of intervening 
processes. 

Comparison of these processes with phase 
transition in classical systems demonstrates their similarities. In classical systems usually an 
external force such as a thermal bath, an ecological pressure~\cite{phtecology}, overload of 
communication networks~\cite{phtnetwork}, etc., which are not controlled or significantly affected 
by the state of the system, pushes it to an attraction point of its phase space irrespective of 
details. This is very similar to the role of environment for quantum systems. In some cases, such 
as a thermal bath for cooling a macroscopic object, the driving force has a single source with a 
clear boundary analogous to a barrier separating a quantum system from its environment. In other 
cases, such as in ecological or artificial networks, the driver is distributed. These cases may 
be compared e.g. with global decoherence of quantum fluctuations due to interaction between 
shorter modes and unobservable long modes extended beyond observer's horizon~\cite{qmdecoherbook}. 
Assuming that the coupling between system and apparatus-environment is switched on at an initial 
time, e.g. the moment an electron enters a Stern-Gerlach detector, the effect of coupling 
influences the state of closest components/degrees of freedom of the apparatus-environment, which 
in turn affect the state of other components. However, weaker interactions, less perturbation 
in the state of the system. With passage of time, one expects that an equilibrium will be reached where the 
state of some components are locked to a state determined by the state of the system. This is when 
entanglement occurs and a state similar to (\ref{sysenventang}) is reached. 

Mathematically, the state of infinite number of components/degrees of freedom of system-environment 
or $A_i$ coefficients of their projection on an arbitrary base $|A_i\rangle$ of the state  
$|\psi, E\rangle = \sum_i^{N \rightarrow \infty} A_i |A_i\rangle$ can be treated as a random field. It 
is proved that under suitable conditions, which should be satisfied by a suitable measurement 
apparatus, the probability distribution function of a random field to which a Hamiltonian is 
associated, is a limit Gibbs distribution, see e.g.~\cite{phtsinai,phtmath} and 
references therein. Assuming that perturbation of environment-apparatus induced by the system is 
small, the Hamiltonian has a number of ground states and distribution of the random field is 
composed of disjoint limit Gibbs distributions around these states. In the case of the random field 
defined by states in (\ref{sysenventang}), before completion of entanglement these disjoint 
distributions around ground states are what leads to formation of equivalence classes described in 
Sec. \ref{sec:largeenv}.

\subsubsection{Sinai  theorem of phase transition} \label{sec:sinai}
The Sinai theorem of phase transition~\cite{phtsinai} proves that for perturbations 
(noise/temperature) lower than a critical value, there always exists a set of small critical 
couplings (smaller than an upper limit) such that each limit Gibbs distribution is in a single phase, 
that is the system has an asymptotic probability of 1 to be in one of the stable ground states. The 
interaction Hamiltonian of system-environment $H$ presents their couplings. Considering 
Hamiltonian of classical systems and their quantum analogues, interaction Hamiltonian of an 
arbitrary system-environment should have the following general form:
\be
H \equiv \sum_{i,j = 1}^{N \rightarrow \infty} f (A_i, A^*_j, \mu_{ij}), \quad \quad 
\langle E,\psi|E,\psi\rangle = \sum_{i,j = 1}^{N \rightarrow \infty} A_i A^*_j = 1 
\label{measurehamilt}
\ee
where $\mu_{ij}$ are couplings. The second equality is a constraint on coefficients when they 
are interpreted as probabilities (see Sec. \ref{sec:random}). Although we have used a countable 
index for system-environment pointer states, they can be uncountable and continuous. In this limit 
cases coefficients $A_i$ can be considered as a random field $A$. Then, Sinai theorem 
shows that for a combination of couplings $\mu_{ij}$ and small perturbation of $A$, the system 
approaches one of stable minima of the Hamiltonian, irrespective of details of 
(\ref{measurehamilt}). 

Considering again a Stern-Gerlach experiment, the environment is composed of 
photons of the magnetic field of the detector and spins of atoms in the detector and outside of it. 
Although all these components contribute to state $|E,\psi\rangle$, only a limited number of 
photons get the chance to interact with the passing electron\footnote{Nonetheless this is usually a 
very large number and Sinai theorem can be applied.}. The interaction Hamiltonian between two 
spins is proportional to $\hat{\vec{S}}_1.\hat{\vec{S}}_2$. Therefore, the classical Hamiltonian 
(\ref{measurehamilt}) depends on $A_iA_j \propto s_i s_j$ where $s$ is the eigen value of projection 
of one spin on the other, and has a discrete support. Even if the state of the electron is 
initially a superposition of two polarizations, according to Sinai theorem, the 
field $A$ approaches randomly to one of two minima of the Hamiltonian - corresponding to alignment 
along or opposite to detector's field - with a probability very close to 1. This corresponds to 
entanglement of the spin of environment with the system (electron) and breaking of state symmetry 
as explained in Sec. \ref{sec:symmbrdeco}. 

According to Sinai theorem weak perturbations of configuration Hamiltonian makes $A_i$'s 
configuration to approach to one of Peierls pure states, which in the case of a quantum system 
corresponds to one of eigen vectors of the observable, with a measure zero difference. 
In statistical mechanics this phenomenon occurs for specific values 
of couplings. Here we have to assume that the critical value of couplings for occurrence of 
$i^{th}$ Peierls pure phase has a probability proportional to $|a_i|^2$. This ensures that in the 
limit of weak interaction it is the randomness of quantum state due to degeneracy (symmetry) 
rather than the apparatus or environment which determines probability of outcomes. Due to 
stability - in mathematical sense - of Peierls pure states, once the symmetry is spontaneously 
broken, it cannot be restored, that is $A_i$ coefficients preserve their value. In quantum 
language system-environment entanglement is stable and needs large perturbations to break it. 

Application of Sinai theorem proves, in a detail-independent manner, that if the setup of a 
measurement is suitable, the state symmetry does break and its occurrence is a direct consequence of 
quantum entanglement. The theorem also predicts the existence of surfaces in coupling space, in 
which multiple phases can coexist. For quantum systems they correspond to cases where 
interaction with environment cannot completely break the symmetry between states. Thus, 
$|\psi_i\rangle$ are not orthogonal and measurement is of type POVM, or an interference 
between states occurs\footnote{A term $A_iA^*_j$ will appear in the constraint 
iff states $|\psi_i\rangle$ and $|\psi_j\rangle$ are not orthogonal. We assume that these states 
are normalized. Moreover, minimum paths are absolute minimums only if decomposition states 
are orthogonal, otherwise they can be decomposed to a set of normal vectors with coefficients 
less than 1.}. 

\subsubsection{Topology change} \label{sec:toplogy}
It is well known that phase transition is related to topological properties of configuration 
space - in the case of quantum systems the composite system-environment state space, 
see~\cite{phttopo,phttopo0} for mathematical aspects and~\cite{phttopo1} for review of topological 
properties involved in phase transition of strongly correlated systems. In symmetry description 
of quantum mechanics the change of at least one of topological properties, namely the topology of 
state space itself is obvious. As a projective complex space, the topology of state 
space is closely related to $(2N+1)$-spheres. After symmetry breaking by measurements it is 
reduced to a point, that is trivial topology by if the measurement sharp, otherwise to 
$(2N'+1)$-spheres with $N' \leq N$ in POVM. 

This is in clear contrast with 
statistical systems in which measurements do not change configuration space and its geometrical 
properties. Due to this topological change phase transition by quantum measurement or decoherence 
can be classified as first order because of discontinuous reduction of representation from an 
$N$-dimensional state space to a trivial 1-dimensional subspace (or a space of smaller dimension in 
POVM), and resolution of apparatus or strength of its interaction with environment can be considered 
as an order parameter. For instance, in Stern-Gerlach experiment if the magnetic field is too weak, 
deflection for two polarization would be too small and they will not be separable. Other examples 
and their interpretation as phase transition are explained in~\cite{symmbrex,symmbrex0}.
Usually there is another threshold which separates a partially decohered system from classical 
regime in which the resolution in not enough to detect any quantum effect and system behaves classically. 
The space between these thresholds can be called the {\it attraction domain} for decoherence.

\subsubsection {Size does not matter} \label{sec:dyntime}
A question arises here: Should a quantum system-environment ensemble have a large number of 
degrees of freedom such that a large chunk of it plays the role of environment and makes 
measurements on the smaller part meaningful ? Note that Sinai theorem can be applied only to systems 
with many degrees of freedom. Here we argue that this is not necessary, and although decoherence 
helps to understand the classical world more easily, it is not a necessary condition. 

Consider a universe consisting of two particles/subsystems with states 
$|\psi_1\rangle$ and$|\psi_2\rangle$. By definition no other substructure/subsystem exists. The only 
possible observation in this small universe is the comparison between states of the two subsystems. 
Thus, the symmetry group is $\mathbb{Z}_2$. The only apparatus available to an observer, for 
instance subsystem 1, is its own state and an interaction Hamiltonian operator: 
$H = \alpha |\psi_1 \rangle \langle \psi_1|$ for an arbitrary $\alpha$. Due to the projective 
property of states $\alpha = 1$ can be assumed without loss of generality. This Hamiltonian, which 
nothing else than density matrix of subsystem 1, preserves $|\psi_1\rangle$. The observer 
can also define a state $|\bar{\psi}_1\rangle$ such that the application of $H$ to it gives null vector 
$|\oslash\rangle$ and $\langle \bar{\psi}_1 | \psi_1 \rangle = 0$. Note that according to no-cloning 
theorem~\cite{qmcloning} subsystem 1 cannot use the unknown state of subsystem 2 as a resource. 

The space generated by ${|\psi_1\rangle, |\bar{\psi}_1\rangle}$ is a representation of $\mathbb{Z}_2$. 
Evidently, any other basis can be chosen, but the observer cannot be aware of it because the interaction 
defined by $H$ compares other states with observer's state and the result (projection) would be the same. 
In particular, $H |\psi_2\rangle \equiv \beta |\psi_1\rangle + \bar{\beta} |\bar{\psi}_1\rangle$ 
should lead to the measurement of $|\beta|$ and determine the probability of similarity of states 
of the two subsystems. However, to measure $\beta$, the observer 1 must distinguish between its 
own state $|\psi_1\rangle$ and $\beta |\psi_1\rangle$ or determine the average outcome 
$\langle \psi_1| H |\psi_2 \rangle$. But because the state space is projective, $|\psi_1\rangle$ 
and $\beta |\psi_1\rangle$ are indistinguishable. Averaging operation is also impossible and after 
the first operation, states of the two subsystems would be either equal or opposite and stay as 
such for ever. Therefore, it would not be possible to repeat the comparison operation unless there 
are multiple universes of the same kind, and an external observer performs averaging between 
outcomes. Therefore, in accord with $\mathbb{Z}_2$ symmetry of this universe, the observer can 
only verify if its state is completely similar or opposite to its neighbour. After the first 
measurement there would not be any superposition and the little universe looks classical.

The above example shows that if the symmetry of a quantum system is abelian, that is it has only 
abelian observables, their simultaneous measurement completely fixes the state once for ever and 
there would not be further evolution. Therefore, a universe composed of such systems would be static 
and without any notion of time. From this observation we conclude that a dynamical universe is composed 
of systems with some non-abelian symmetries and interactions depending on noncommuting 
observables. In this way a system cannot continue to be in an eigen state of all observables after 
a measurement operation. In this context, position operator as an observable has a special 
importance for dynamical behaviour of systems\footnote{This argument applies to any operator 
$\hat{X}$ and its variation $\hat{D}_X$ such that $\hat{D}_X |\psi (x)\rangle = \partial \psi (x)/ 
\partial x |\psi (x)\rangle $}. By definition in absence of interaction a system which is in an 
eigen state of position is static and does not evolve. In classical physics position and 
momentum or variation of position are independent observables. 
In quantum mechanics the momentum - the generator of translation symmetry - is proportional to 
$i\partial /\partial x$ and does not commute with position operator. Indeed it was meaningless if 
an object could be at the same time in eigen state of position and its variation. This gives a 
logical insight into Heisenberg uncertainty principle and quantum Zeno effect~\cite{zeno}.

The subject of variational operators reminds the issue of time in quantum mechanics. The problem 
arises when diffeomorphism invariant models are quantized. In such models the total Hamiltonian, 
which generates time variation in quantum systems, is null. This means that the Universe as a whole 
is static. As we discussed earlier, in the symmetry description of quantum mechanics a single system without 
subsystems is equivalent to trivial representation of symmetries. Therefore, if the state of 
entities in the Universe are neglected, its state cannot be anything else than a static and trivial 
representation of any symmetry group. Thus, only relative evolution of subsystems in an intertwined 
Universe has a physical meaning. This conclusion is consistent with relative probability and 
{\it internal clock} proposed by Page and Wootters~\cite{timepwmodel} and its 
modification~\cite{timepwmodelmod} to resolve problems regarding the absence of transition and 
propagation in the original formulation~\cite{timepwcrit}. 

\subsubsection{Physical reality of state space and nonlocality} \label{sec:measurestate}
The definitive modification of the state space after a measurement is usually considered to put 
doubt in the status of the state as a physical property of the system, even when the inverse is 
proved by the PBR no-go theorem~\cite{qmnooverlap}. The description of quantum mechanics in symmetry 
language shows that the concept of reality of the state space and not being measurable or invariant 
are not mutually exclusive. In fact, even in macroscopic world we can find many analogous examples. 
For instance, in set theory, if no subset is defined, all members have exactly the same relation 
with respect to the set and are permutable. But if we define two subsets, although the nature of 
the set and its members is not modified, they acquire a difference and an additional structure, because 
they can belong to one or other subset and permutation symmetry is partially broken. If the state of an 
element is defined as its membership of one or other subset, before an attempt to find out this attribute, 
its state can be defined as a {\it classical} superposition of two subsets. But by definition its measurement 
corresponds to breaking this degeneracy. Therefore, it is logically wrong to say that state is not a physical 
reality because it changes when measured.

Describing quantum mechanics in symmetry view also shed light on nonlocal entanglement. Depending 
on the type of interaction between subsystems some of their properties may be entangled where 
others stay independent because the symmetry group can be direct product of lower rank groups. 
In particular, in contrast to classical physics spacetime does not have any special role and it 
can be orthogonal to other symmetries, leading to nonlocal entanglement of the state and 
observables. An important conclusion of this observation is that systems are only approximately 
isolated and independent. Considering the universality of gravitational interaction, this 
conclusion may be turned around: It may be the indivisibility of the Universe which reflects 
itself as a universal attractive force that we call gravity. We briefly review some ideas in this 
direction in Sec. \ref{sec:qmgr}.

\section{Fundamental randomness and probability} \label{sec:random}
Axiom (\ref{postsymmbr}) is very similar to its analogue in standard quantum mechanics, but it does 
not specify how the probability is determined, because we obtain it in this section from quantum mechanics axioms 
and some general properties of probability distributions. 

Since the discovery of quantum mechanics and observation of randomness in outcome of 
measurements - unsharp outcomes~\cite{qmenercons}, physicists have put their hope on 
{\it hidden variables}~\cite{hiddenvar}, coarse-graining of spacetime~\cite{consistenthist}, 
and many other means to establish {\it hidden rules} and explain the randomness of 
quantum systems in the same line of reasoning as in classical physics. However, 
experimental verification of Bell's inequalities~\cite{bellinequalver0,bellinequalver1,
bellinequalver2} and other quantitative tests~\cite{eprtimevar,qmnooverlap} 
demonstrate the contrary. 

Following axiom (\ref{poststate}), the state of a system contains all obtainable information about 
it, including probabilities associated to pointer states consisting of eigen states of independent 
observables. Because any vector belonging to the state space can be decomposed to a pointer basis, 
it is logical to conclude that the expansion coefficients must be related to the probability that 
the system be in the corresponding eigen state when the symmetry/degeneracy is broken. Giving the 
fact that probabilities must be positive real numbers, the square of the absolute value of 
coefficients $|a_i|^2$ is the most natural candidate. As state space is projective, without loss 
of generality we can assume that pointer states are normalized. Thus:
\be
\sum_i^n |a_i|^2 = 1 \label{coeffsum}
\ee
where $n$ is the dimension of the state space and can be $\infty$ or even innumerable. Equation 
(\ref{coeffsum}) is a necessary condition for interpreting $|a_i|^2$ as probabilities. However, 
apriori any positive real function $f_i (|a_i|^2)$ can be an equally valid candidate. In this case 
normalization of the probability leads to:
\be
\sum_i^n f_i (|a_i|^2) = 1 \label{probsum}
\ee
Consider two copies of the same system prepared independently but in the same manner. According to 
postulate \ref {postcomposite} the state space of the composite system made from these subsystems is:
\be
|\psi\rangle = \frac{1}{2}(|\psi_1 \rangle \otimes |\psi_2\rangle + |\psi_2 \rangle \otimes 
|\psi_1\rangle) = \sum_i a_i^2 |i\rangle \otimes |i\rangle + \sum_{i \neq j} a_i a_j 
(|i\rangle \otimes |j\rangle + |j\rangle \otimes |i\rangle) \label{compsys}
\ee
Note that even for entangled or interacting subsystems the state of the composite system can be written as l.h.s. 
of (\ref{compsys}), but some of coefficients $a_i$ may be zero. Here symmetrization over two subsystems means
that due to similar preparation they are indistinguishable. 

From \ref{compsys} we conclude that the probability for subsystems to be in $(i,j)$ state is 
$f(|a_ia_j|^2)$. On the other hand, because these subsystems are prepared independently, the mutual 
probability for the system to be in state $(i,j)$ is $f(|a_i|^2)f(|a_j|^2)$. Therefore:
\be
f(|a_i|^2)f(|a_j|^2) = f(|a_i|^2|a_j|^2) \label{mutualprob}
\ee
It has been proved~\cite{qmhardy} that the only function with such property is a positive 
power-law\footnote{More exactly, in~\cite{qmhardy} this theorem is proved for the case in which 
the function $f$ is applied to positive integer numbers. Isomorphy of integer and fractional numbers 
extends (\ref{mutualprob}) to rational numbers. However, because fractional numbers are a dense subset of 
real numbers, the power-law form must be the unique solution for all real numbers as long as we 
assume that $f$ is a smooth function with at most countable number of discontinuities.}. Thus:
\be
f(x) = x^\beta~,~\beta > 0 \label{fform}
\ee
Considering this property along with (\ref{coeffsum}) and (\ref{probsum}), it is straightforward 
to see that only $\beta = 1$ can satisfy all these relations. Therefore, $f (|a_i|^2) =|a_i|^2$ is 
the only possible expression for the probability of pointer states. When the basis is normalized, 
coefficients $|a_i|^2$ form a $(n-1)$ simplex similar to classical systems. We remind that  
no condition on the dimension of the configuration space is imposed in the above argument. 
Thus, when $n \longrightarrow \infty$, equations (\ref{probsum}) and (\ref{fform}) impose $L^2$ 
integrability condition on the state space and it must be a Hilbert space. This completes the 
proof that axioms presented in Sec. \ref{sec:qmaxioms} leads to standard quantum mechanics, 
and from this point there is no difference between the latter and the construction of quantum 
mechanics according to axioms (\ref{poststate}) to (\ref{postsymmbr}) except that these 
axioms relate the foundation of quantum mechanics to symmetries.

Here we obtained Born rule for probability of sharp measurements outcome for pure states. 
Because mixed states have incomplete symmetry breaking density matrix, they can be virtually 
completed to pure states. Therefore, the probability rule discussed here apply to them too.

\section{Dynamics, decoherence and classicality in symmetry description} \label{sec:decohere}
Axiom (\ref{postunitary}) is the same as its analogue in standard quantum mechanics except that it 
explicitly insists on the role of representation of symmetries realized by the state space of the 
system. An immediate consequence of this postulate is a difference between dynamics equation for 
bosons and fermions which belong to different representations of Lorentz group, namely vector and 
spinor representations, respectively. As we discussed in the introduction, several issues regarding 
evolution of quantum systems such as existence of interference; and decoherence and apparently 
nonunitary evolution of state after sharp measurements are still considered to be confusing.
In this section we briefly review these topics with regard to what the introduction of symmetry as a 
foundational concept in quantum mechanics can do for their clarification.

\subsection{Decoherence and classical behaviour} \label{sec:decoherecalssic}
We discussed decoherence in the framework of measurement in Sec. \ref{sec:measure} and 
argued that it is a proper replacement for collapse. However, a number of issues 
about decoherence are raised in the literature. They can be summarized as the 
followings~\cite{qmphil,decohererev2,decohererev3}:
\renewcommand{\theenumi}{\alph{enumi}}
\begin{enumerate}
\item Interference is only partially removed by decoherence. \label{decohinterf}
\item Decoherence needs open systems. Thus, it cannot be applied to the Universe as a whole. 
\label{decohopen}
\item In many quantum phenomena the division into system an environment is somehow arbitrary. 
In these cases how does this arbitrary separation influence the state to which the system 
decohere ?  \label{decohsep}
\item Decoherence in relativistic quantum field theory. \label{decohqft}
\end {enumerate}
To these criticisms we must add the ambiguity of pointer basis which we discussed in 
Sec. \ref{sec:measure}. We also addressed system-environment separation. Here we 
concentrate on the remaining issues and discuss how symmetry description may 
help to understand them.

\subsubsection{Suppression of interference} \label{sec:interfersupp}
In the Introduction section we suggested a bifurcation (\ref{unitarybifur}) as a counter example 
to criticism about the inevitability of non-unitary evolution of state during 
decoherence~\cite{qmdecoherprob}. The foundational presence of symmetries implies that in 
macroscopic apparatus-environment with large number of degrees of freedom, symmetries are continuous. 
Then, the Sinai theorem on the presence of solution flows and how a dynamical system approaches 
randomly to one of them and breaks the symmetry, provides a mathematical support for 
existence of behaviour similar to (\ref{unitarybifur}). In addition, the theorem about decoherence 
in absence of common basis or reference frame between system-apparatus and environment, in which 
symmetries play a central role, proves that symmetry breaking and decoherence occur independent of 
how big or small are these subsystems.

\subsubsection{Global decoherence} \label{sec:globuniv}
In what concerns classicality of the Universe as a whole, in Sec. \ref{sec:insepar} we argued that 
Universe must be considered as an intertwined ensemble of its components and its global state 
is simply union of the states of its components. If the components are decohered, so is the Universe. 
Nonetheless, very large anisotropy modes of the Universe may yet have a coherent state, because their 
causal interaction is suppressed by the expansion of the Universe~\cite{houricond}. Indeed, 
simulation of quantum corrections in a toy model of early Universe shows that nonlocal quantum 
effects may have crucial contribution in both inflation and late acceleration of the 
Universe~\cite{houricondqm}. Giving the fact that at present dark energy is the dominant constituent 
of the Universe, these observations may mean that only a fraction of the Universe behaves 
classically. 

The example of a small universe in Sec. \ref{sec:dyntime} shows that a probabilistic interpretation 
of measurements performed on large modes of matter distribution in the Universe may be meaningless, 
because for them Universe is {\it small} and they have access only to a single copy of it. The 
existence of these causally inaccessible modes makes the Universe effectively an open system. 
Nonetheless, the influence of processes occurring outside the past and possibly future horizons on 
everything inside are very small and negligible. Therefore, for all practical (local) applications 
the Universe behaves classically.

\subsubsection{Effect of system-environment separation on measurements} \label{sec:sysenvsep}
Regarding (\ref{decohsep}), apriori it seems that definitions of system and environment and what is 
what influence the outcomes. This is certainly a fact. For instance, in the example of a decaying pion in 
Sec. \ref{sec:state}, considering a magnetic field, even a weak one, as part of the system completely 
changes the definition of the system from first place, and thereby symmetries that define the state 
space, possible outcomes, etc. The situation would be different if we change the environment. 
According to axiom (\ref{poststate}) a system is defined by its symmetries. Therefore, 
characteristics of the environment do not matter for the system as long as they do not change its 
symmetries. In the example above, including the magnetic field in the environment does change 
its interaction with the system, but an observer would not be able to conclude its presence from 
measuring the probability of unknown spin state of the system.

\subsubsection{Decoherence in Quantum Field Theory (QFT)} \label{sec:qftdecoh}
Finally, (\ref{decohqft}) refers to the claim that relativistic quantum field theories do not admit a 
superselection~\cite{qmmeasuresupersel}, and thereby algebraic description cannot explain 
decoherence. However, the entanglement between system and environment defined by 
(\ref{sysenventang}) is valid for quantum fields, and the theorem about decoherence in absence 
of common reference frame applies too. Although the notion of environment in QFT is 
loose, for perturbative systems far from interaction zone, particles can be considered as isolated 
and usual rules of decoherence discussed in Sec. \ref{sec:measuredecoher} are applicable, see 
e.g.~\cite{qmdecoherbook} for detailed discussion and examples of decoherence of relativistic 
particles.

As for non-perturbative systems, they are in highly correlated/entangled states and far from 
classicality. Their strong interactions and quantum correlations by symmetries provide a natural 
pointer basis for these systems. Due to extreme physical conditions for their formation and 
despite their theoretically infinite dimension, non-perturbative systems are very sensitive to 
environment and decoherence~\cite{qftdecoher}. It is worth to remind that in these systems 
symmetries play a crucial role and strong quantum correlation and entanglement of degrees 
of freedom lead to breaking of symmetries and a wealth of nonlocal and collective quantum 
effects in condensed matter. 

\section{Gravity in a quantum universe} \label {sec:qmgr}
Relation between gravity and quantum mechanics or in short {\it Quantum Gravity} 
is a vast topic by its own and cannot be properly reviewed here. Therefore, in this section we 
briefly review a couple of works and ideas in progress which may be useful in search for a 
consistent description of gravitational phenomena in the realm of quantum mechanics. 

\subsection{Analogies and universalities}
In contrast to many approaches to quantum gravity which try to quantize gravity field or 
geometry in one way or another, here we seek an inherent relation between the two models. 

The idea of an intrinsic relation between gravity 
and quantum mechanics is not new. The first evidence may be the black hole entropy and 
its analogy with thermodynamics~\cite{hawkingrad,entropylimit}  However, it is well known 
that  black hole entropy is a purely geometrical property. It is the Norther charge of 
diffeomorphism symmetry and independent of the special case of Einstein 
gravity ~\cite{waldnoether,waldpapers}. More recently AdS/CFT correspondence - an 
isomorphism between gravity in a bulk spacetime with  AdS geometry and conformal field 
theories on the boundary surface~\cite{adscft} - has raised hope that QFT models, notably 
gauge theories can be related to geometry. However, there is not yet a general consensus 
how quantum operators on the boundary should be projected to the bulk. Nonetheless, it 
has been shown that quantum entanglement on in the CFT model and its decreasing can be 
interpreted as separation of subspaces in the AdS bulk~\cite{adsentangle}. 

None of the above analogies provides a physical insight into the role of gravity in a quantum 
world. However, it seems that without gravity, quantum mechanics is an incomplete theory 
because it does not provide any mass or distance scale~\cite{houriqgr}. This issue is related 
to another question:

Why is the Planck constant $\hbar$ universal ?

We remind that $\hbar$ does not appear in the axioms of quantum mechanics and its presence 
becomes necessary only in the definition of momentum and Heisenberg uncertainty relation. 
However, apriori this constant which determine the degree of correlation between position and its 
variation may be different for different systems. The universality of $\hbar$ means that the 
amount of randomness in the dynamics of physical systems is the same regardless of their 
mass, size, and couplings. If the value of $\hbar$ were not universal, we could parametrize 
it by a factor which we call $\alpha$. Then, Schr\"odinger-Klein-Gordon equation becomes:
\be
(\alpha^2\hbar^2 \Box - m^2) |\psi\rangle = 0 
\Longleftrightarrow 
(\hbar^2 \Box - m'^2) |\psi\rangle = 0~,~~~~~ \quad m' = \frac{m}{\alpha}~. \label{schrod}
\ee
Therefore the non-universality of $\hbar$ can be removed by redefinition of mass. However, mass is 
the gravity charge ! In the same way a different coupling $G$ to gravity can be removed by redefinition 
of mass which in its turn modifies the Schr\"odinger equation and can be presented as a scaled $\hbar$.
Even if mass is generated by interaction, the scaling can be performed on the vacuum expectation value. 
Therefore, the universality of gravity and dynamical randomness in quantum mechanics are related.

\subsection{Ambiguity of a classical universal force in a quantic Universe}
Inconsistency of a classical gravity in a quantum world is well known~\cite{qmgrinconsist}. Here we present 
this issue with a simple example: \\
Consider a large empty box/potential well in Minkowski space. A single particle of mass $m$ and precisely 
known energy and momentum is thrown in the box\footnote{To remove the ambiguity due to black hole formation, we 
consider a small but finite size for the particle larger than its Schwarzschild radius.}.Because energy and momentum 
of the particle are known, according to Heisenberg uncertainty principle all information about its location inside the box 
is lost and its wave function is uniform up to a phase $exp(ikx)$, where $k$ and $x$ are 4-vectors of momentum and 
position, respectively. This means that the spacetime remains homogeneous and the presence of a massive particle 
does not breaks the translation symmetry. The semi-classical Einstein equation with a constant energy-momentum 
tensor:
\be
R_{\mu\nu} -\frac{1}{2} g_{\mu\nu} R = \frac{8\pi G}{c^4}\langle\psi|\hat{T}_{\mu\nu}|\psi
\rangle \sim \eta_{\mu\nu} \Lambda~, \quad\quad \Lambda \rightarrow 0~. 
\label {desitter}
\ee
leads to a De Sitter metric which is conformally flat and as $\Lambda = m/V \rightarrow 0$, where $V$ is the 
volume of the box, the geometry approaches to a Minkowski space. By contrast, a classical particle in a flat 
Minkowski space changes the symmetry to spherical (in the rest from of the particle) and changes the flat 
metric to Schwarzschild. If a second particle with the same conditions for its energy-momentum is added to the 
box, the solution of Schr\"odinger equation will depend on the distance of particles. But as their initial positions
are unknown, one has to integrate over all possible values. Thus, for an observer their wave function would 
be again homogeneous and the observer concludes that they do not {feel} each other presence through gravitational 
attraction !

\subsection{Gravity as minimum entanglement}
In Sec. \ref{sec:measurestate} the universality of gravitational interaction was used to argue that all 
subsystems of the Universe are somehow correlated/entangled through their gravitational coupling. 
This observation motivates a purely quantum mechanical definition for gravity through a slight 
modification of postulate \ref{postcomposite}~\cite{houriqgr}:

The homomorphism which relates the Hilbert space of a composite system to those of its components 
never becomes an isomorphism. The minimum interaction/entanglement is perceived as gravity and in the 
limit of macroscopic decohered systems it is ruled by Einstein equation.

This axiom can be formulated as the following:
\be
{\mathcal H} \neq {\mathcal H}_1 \otimes {\mathcal H}_2, \quad 
{\mathcal H} : {\mathcal H}_1 \otimes {\mathcal H}_2 
\xrightarrow{\mbox{\footnotesize gravity}} 
{\mathcal H}_1 \otimes {\mathcal H}_2 ~\biggl | ~x \in {\mathcal H}_1 \otimes 
{\mathcal H}_2~,~y = {\mathcal H} (x) \in {\mathcal H}_1 \otimes {\mathcal H}_2 \label{proj}
\ee
According to this definition a single fundamental particle with known energy and momentum {\it does not feel} 
its own gravity! A direct consequence of this property is that such a particle does not make a 
black hole. By contrast, if the same particle is localized, then its energy-momentum would be 
completely uncertain similar to particles inside a black hole - for a far observer can be at any 
place inside the horizon. Thus, similar to entanglement in standard quantum mechanics, gravity 
exists only if there are at least two particles/subsystems/degrees of freedom.

\subsection{Spacetime and dynamics}
Definition of gravity through entanglement does not specify how it is determined and related to 
other observables and symmetries. Nonetheless, we expect that in the classical limit the dynamic 
equation approaches the Dewitt-Wheeler equation with minimal interaction to gravity of environment:
\be
\hbar^2 \Box + \frac{R}{6}- m^2)|\psi (x)\rangle = 0~.  \label{schrodKK}
\ee
Moreover, we know that in classical limit it is always possible to find a local inertial frame in which 
particles do not {\it feel} the gravity and are free. Thus, we consider the following transformation from one-particle 
coordinates $x$ to {\it free} coordinates $X$:
\be
\hat{X}_i = \hat{x}_i \quad , \quad \hat{P}_i = \hat{p}_i + \sum_{j \neq i} 
f (\hat{x}_j, \hat{p}_j)~,  \label{deffree}
\ee
where $f (\hat{x}_j, \hat{p}_j)$ is an unspecified function. It is easy to show that commutators become:
\bea
&& [\hat{x}_i , \hat{x}_j] = [\hat{X}_i , \hat{X}_j] = 0 \quad i,j = 1, 2, 
\ldots \label{xcommu} \\
&& [\hat{p}_i , \hat{p}_j] = [\hat{P}_i , \hat{P}_j] = 0 \label{pcommu} \\
&& [\hat{x}_i , \hat{p}_i] = i\hbar \label {xpicommu} \\
&& [\hat{X}_i , \hat{P}_i] = i\hbar \label {xppicommu} \\
&& [\hat{x}_i , \hat{p}_j] = [\hat{X}_i , \hat{P}_j] = 0 \quad i \neq j 
\label {xpijcommu} \\
&& [\hat{x}_i , \hat{P}_j] = [\hat{x}_i, f (\hat{x}_i , \hat{p}_i)] 
\label {xppijcommu}
\eea
In equations (\ref{xcommu})-(\ref{xppijcommu}) the indices refer to particles and the spacetime 
indices are neglected. We assume the following form for the function $f$:
\be
f (\hat{x}_i, \hat{p}_i) = \frac{1}{\Lambda\hat{x}_i}\hat{p}_i \label {fexpan}
\ee
where $\Lambda$ is an energy scale. Then:
\bea
&& \hat{P}_i = \hat{p}_i + \sum_{j \neq i} \frac{1}{\Lambda\hat{x}_j}\hat{p}_j~, \label{ppi} \\
&& [\hat{x}_i , \hat{P}_j] = [\hat{X}_i , \hat{P}_j] = \frac{i\hbar}
{\Lambda \hat{x}_i} \label {xppfcommu}~.
\eea
Finally, we apply (\ref{ppi}) to the Schr\"odinger-Klein-Gordon equation (\ref{schrod}):
\be
\biggl\{\sum_i \hat{p}_i^2 + 2 \sum_{j \neq i} \frac{1}{\Lambda \hat{x}_j} \hat{p}_i\hat{p}_j
 + \sum_{j,k \neq i} \frac{1}{\Lambda^2 \hat{x}_j\hat{x}_k} \hat{p}_j\hat{p}_k + m^2 \biggr \}|\psi\rangle = 0~. \label{schrodgravity}
\ee
The operator part of equation (\ref{schrodgravity}) has the same structure as (\ref{schrodKK}) 
and includes terms which can be interpreted as gravitational interaction. Therefore, there is hope 
that the extension of this formulation to a continuum leads to a model which at classical limit look like Einstein gravity.

\subsubsection{Unifying spacetime with its content}
In the formulation above we neglected spacetime indices. Aside from simplifying the notation, 
there is a deeper reason for this ignorance. First of all it proves that this formalism can be 
applied to any background spacetime. More importantly, the similarity of the role of 
species/particle index to spacetime indices indicates that there is an interchangeable role 
between what is called {\it a particle} and {\it a point in the spacetime}. Therefore, we can claim 
that in this model there is a natural unification or embedding of the spacetime with a group 
manifold determining the symmetries and variety of particles. 

Motivated by the above observation and the foundational role of symmetries, we can abstract 
this model further\footnote{This is a work in progress and here we presents preliminary ideas.}:

Consider a set of $N \rightarrow \infty$ entities\footnote{Ultimately we want to extend $N$ to a 
continuous cardinal number.}. We assume that physical processes are invariant under permutation 
group $\Pi(N)$ of these entities. The Hilbert space $\mathcal{H}$ representing this quantum system 
has an extended $SU(N), N \rightarrow \infty$ symmetry. Large groups, including $SU(N)$, are in 
general decomposable to direct products of smaller groups. Therefore, we assume that 
$SU(N)\biggl |_{N \rightarrow \infty} = G_0 \otimes G$ where the rank of $G_0 \rightarrow \infty$ 
and $G$ has a finite rank. Here we concentrate on $G_0$. As a compact Lie group operator 
$\hat{X}_i \in \mathcal{B}[\mathcal{H}]$ satisfy commutation relations:
\be
[\hat{X}_i ,\hat{X}_j] = L~f_{ijk} \hat{X}_k, \quad i,j,k \in {1,2,\dots,\infty} \label{commytsoace}
\ee
where $L$ is a constant. If we assume a dimensionality for $\hat{X}$ operators, $L$ should 
have the same dimension. Moreover, the extension of permutation symmetry of the entities 
to a continuous $SU(N)$ symmetry was owed to quantization. Therefore, we expect that 
coefficients $f_{ijk}$ be proportional to $\hbar$. Thus, in the limit of $L \rightarrow 0$ or 
$\hbar \rightarrow 0$, operators $\hat{X}_i$ commute. Moreover, for $N\rightarrow \infty$ 
the set of $\hat{X}_i$ is isomorphic to real numbers and the set of commutation relations 
(\ref{commytsoace}) can be written as:
\be
[\hat{X},\hat{Y}] = L~f(x,y,z)\hat{Z} \label{commutxyz}
\ee
where the 3 operators in (\ref{commutxyz}) are invariant under a rotation $SO(3)$ group. 
In addition, the latter symmetry is independent of the value of $\hbar$. Therefore, it has 
all the necessary properties to be considered as symmetry of 3 axes of the real space. 
This simple construction shows why we probably perceive the Universe as a 3 dimensional 
space because of its huge degrees of freedom and the structure of Lie groups.

This model is very far from being complete and investigation of the properties of coefficients 
$f(x,y,z)$, definition of a time coordinate, and how a classical gravity may emerge from this 
fundamentally quantum construction is under investigation.

\section{Outline} \label{sec:summary}
Symmetry is the backbone of logical perception of the Universe. In this review we showed that 
when axioms of quantum mechanics, which is the most fundamental description of physical world, is 
formulated such that the role of symmetries is highlighted, puzzles and mysteries of quantum 
mechanics are better explained and clarified. Along and in accordance to this point of view, we 
argued that gravity may have more intimate relation with quantum nature of the Universe than our 
attempts for finding a quantum gravity theory have presumed so far. We described preliminary ideas 
about how gravity may be included in quantum theories. However, the task is very far from accomplishment
and achievement of the goal not easy. Until then, efforts must continue and new ideas welcomed.

\section*{Acknowledgments}
The author would like to thank G.E. Volovik for bringing to her attention Sinai theorem on phase 
transition and its application in quantum mechanics.

\end{document}